\documentclass[twocolumn,showpacs,preprintnumbers,amsmath,amssymb,aps,prl]{revtex4-2}
\usepackage{graphicx}
\usepackage{dcolumn}
\usepackage{bm}
\usepackage{amsmath,amssymb}
\usepackage{epstopdf}
\usepackage{psfrag}
\usepackage[T1]{fontenc}
\usepackage{hyperref}
\usepackage{tabularx}
\usepackage{color}
\usepackage{float}

\newcommand\numberthis{\addtocounter{equation}{1}\tag{\theequation}}

\newcolumntype{Y}{>{\centering\arraybackslash}X}
\begin{document}

\preprint{}

\title{Molecular dynamics simulation of the ferroelectric phase transition in GeTe: displacive or order-disorder?}
\author{{\DJ}or{\dj}e Dangi{\'c}\textsuperscript{1,2}}
\email{dorde.dangic@ehu.eus}
\author{Stephen Fahy\textsuperscript{1,2}}
\author{Ivana Savi\'c\textsuperscript{2}}
\email{ivana.savic@tyndall.ie}
\affiliation{\textsuperscript{\normalfont{1}}Department of Physics, University College Cork, College Road, Cork, Ireland}
\affiliation{\textsuperscript{\normalfont{2}}Tyndall National Institute, Dyke Parade, Cork, Ireland}

\date{\today}

\begin{abstract}
Experimental investigations of the phase transition in GeTe provide contradictory conclusions regarding the nature of the phase transition. Considering growing interest in technological applications of GeTe, settling these disputes is of great importance. To that end,  we present a molecular dynamics study of the structural phase transition in GeTe using a machine-learned interatomic potential with ab-initio accuracy. 
First, we calculate the asymmetric shape of the radial distribution function of the nearest-neighbor bonds above the critical temperature, in agreement with previous studies. However, we show that this effect is not necessarily linked with the order-disorder phase transition and can occur as a result of large anharmonicity. Next, we study in detail the static and dynamic properties of the order parameter in the vicinity of the phase transition and find fingerprints of both order-disorder and displacive phase transition. 

\end{abstract}


\maketitle

\section{I. Introduction} 

The phase transition in ferroelectric materials is usually discussed in terms of two distinct mechanisms, which determine whether the phase transition has order-disorder or displacive character \cite{STEIGMEIER, main, newmain, displejsiv, Fons, macunaga}. The distinction between these two mechanisms comes from the analysis of a simplified Landau model of ferroelectric materials \cite{landau}. 
In the displacive limit of the phase transition, the frequency of a soft phonon mode becomes zero in the higher symmetry structure at the critical temperature. The soft phonon mode freezes in the lower symmetry structure driving the structural phase transition~\cite{STEIGMEIER, main, newmain, displejsiv}. On the other hand, in the order-disorder limit of the phase transition, the local ferroelectric distortion persists above the critical temperature. In this case the paraelectric nature of the high-symmetry phase stems from the lack of the long-range spatial correlation of the polarization \cite{Fons, macunaga}.

Germanium telluride, GeTe, is an important thermoelectric material that is also ferroelectric below 600 - 700 K \cite{TCnote,levin, Wu2017, pei-joule-gete, biswas-gete-rev}. The Landau model of ferroelectric phase transitions places GeTe at the boundary between materials exhibiting order-disorder and displacive characters of the ferroelectric phase transition \cite{BrucePT}. This is further confirmed by a number of experimental studies with contradictory conclusions \cite{STEIGMEIER, main, newmain, displejsiv, Fons, macunaga}. Depending on the spatial resolution of the experimental method, the phase transition in GeTe is found to be either order-disorder or displacive. This ambiguity suggests that a computational, first-principles based study of the phase transition would provide useful insights. 

Our recent works have been able to explain a number of interesting properties of GeTe at the ferroelectric phase transition, primarily  negative thermal expansion \cite{OurNTE} and an increase of the lattice thermal conductivity \cite{OurTC}. However, both of these studies relied heavily on a phonon picture of GeTe, implying a displacive character of the phase transition. There is an open question of whether the inclusion of order-disorder character in the calculations would lead to different results and conclusions.

Molecular dynamics (MD) simulations are probably the most direct tool for classical simulations of materials \cite{IvanaMD, GalliMD, PbTeMD}. In principle, they can capture all relevant physical effects at high temperatures, where quantum corrections are negligible. However, MD simulations for systems containing many atoms when forces are determined by density functional theory (DFT) are extremely computationally expensive~\cite{ChatterjiMD, GeTeAIMD}. To circumvent this issue, researchers usually rely on a simple analytic form of interatomic potentials which have limited accuracy and transferability \cite{Tersoff, RuanBiTe, JavierMD}. Recent works on machine learning interatomic potentials aim to correct this and provide interaction models of similar quality to DFT, at a much more modest computational price \cite{GAP1, GAP2, NN1, descriptor1, Bosoni2019}. These interatomic potentials have been recently used to describe phase transitions in a variety of materials \cite{GeTeNN1, GeTeNN2, MocanuGeTe, Wang2021, Verdi2021}. 

In this paper, we present a molecular dynamics study of the ferroelectric phase transition in germanium telluride. To calculate atomic forces and energies along MD trajectories, we used our recently developed interatomic potential for GeTe using the Gaussian Approximation Potential (GAP) framework \cite{GAP1, GAP2}. Our model of interatomic interactions in GeTe, based on DFT energies and atomic forces, reproduces the experimental structural parameters and negative thermal expansion at the phase transition. The radial distribution function of the nearest-neighbor bonds in GeTe was found to be strongly non-Gaussian even at temperatures above the phase transition. We show that this does not necessarily mean that the phase transition has an order-disorder character and that this effect could arise as a consequence of strong anharmonicity. Furthermore, we present a detailed investigation of the order parameter behavior at the ferroelectric phase transition, which is found to exhibit fingerprints of both order-disorder and displacive character.

\section{Computational details}

Molecular dynamics simulations were performed using the LAMMPS code \cite{lammps}. To calculate the atomic forces and energies, we used our previously fitted Gaussian Approximation Potential (GAP) for GeTe~\citep{OurTC, GapGeTe}. This interatomic potential was fitted to DFT energies and atomic forces. More details about the fitting procedure and the potential are given in Ref.~\cite{OurTC}. To obtain the equilibrium values of structural parameters at different temperatures, we first run a 10 ps simulation using the NVT ensemble to equilibrate velocities at the given temperature, followed with a 20 ps NPT simulation to equilibrate the structure \cite{NPT, NVTNPT}. We then run a 200 ps NPT simulation, while collecting data every 0.1 ps. The timestep is taken to be 1 fs. For convergence study with the size of the simulation region, see Supplementary material \cite{supp}\nocite{NTEPT1, NTEPT2, NTEPT3}. 
At each temperature we start from the zero temperature equilibrium structure of GeTe and set random initial atomic velocities sampled from the normal distribution with the variance corresponding to the target temperature.

To compute the order parameter at different temperatures, we perform 300 ps NVT simulations on a 512 atoms cell while collecting atomic positions every second time step. A time step of 1 fs was used in all simulations. Prior to data collection, we equilibrate the system for 50 ps in the NVT ensemble.

\section{Structural parameters and thermal expansion of GeTe}

\begin{figure}
\begin{center}
\includegraphics[width=0.9\linewidth]{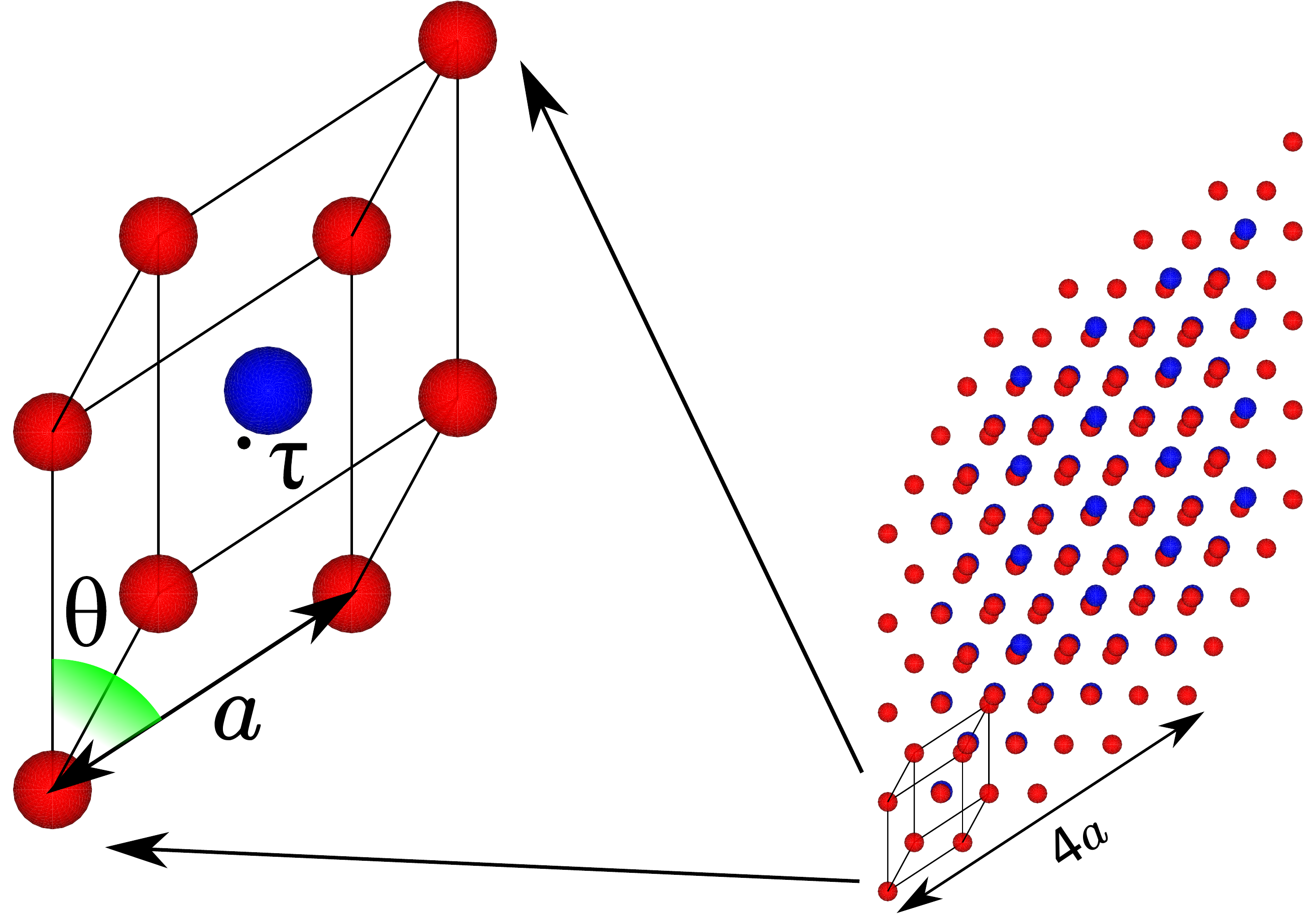}
\caption{Primitive unit cell of GeTe. Red/blue spheres are germanium/tellurium atoms. $a$ is the lattice constant, $\theta$ is the rhombohedral angle, and $\tau$ is the order parameter (the vector from the center of the unit cell, black point, to the tellurium atom). The side image shows the simulation cell in molecular dynamics simulations. For presentation purposes we show the 4$\times$4$\times$4 supercell, instead of the 10$\times$10$\times$10 used in our calculations. The image was made using the Vesta software~\cite{Vesta}.}
\label{fig0}
\end{center}
\end{figure}

Germanium telluride crystalizes in a rhombohedral structure below 600 K (see Fig.~\ref{fig0}), which is described by the following lattice vectors:
\begin{align*}
\vec{R}_{1} &= a(b,0,c), \\
\vec{R}_{2} &= a(-\frac{b}{2},\frac{b\sqrt{3}}{2},c), \numberthis \\
\vec{R}_{3} &= a(-\frac{b}{2},-\frac{b\sqrt{3}}{2},c).
\end{align*}
Here $a$ is the lattice constant of the primitive unit cell of GeTe, $b = \sqrt{\frac{2}{3}(1-\cos\theta)}$ and $c = \sqrt{\frac{1}{2}(1+2\cos\theta)}$. $\theta$ is the angle between the lattice vectors and can be regarded as a secondary order parameter, since in the cubic phase it has a fixed value of $60^{\circ}$, and a lower, temperature dependent, value in the rhombohedral phase. The atomic positions in reduced coordinates are taken to be: Ge (0,0,0) and Te (0.5 + $\tau , 0.5 + \tau , 0.5 + \tau$).

\begin{figure}
\begin{center}
\includegraphics[width=0.9\linewidth]{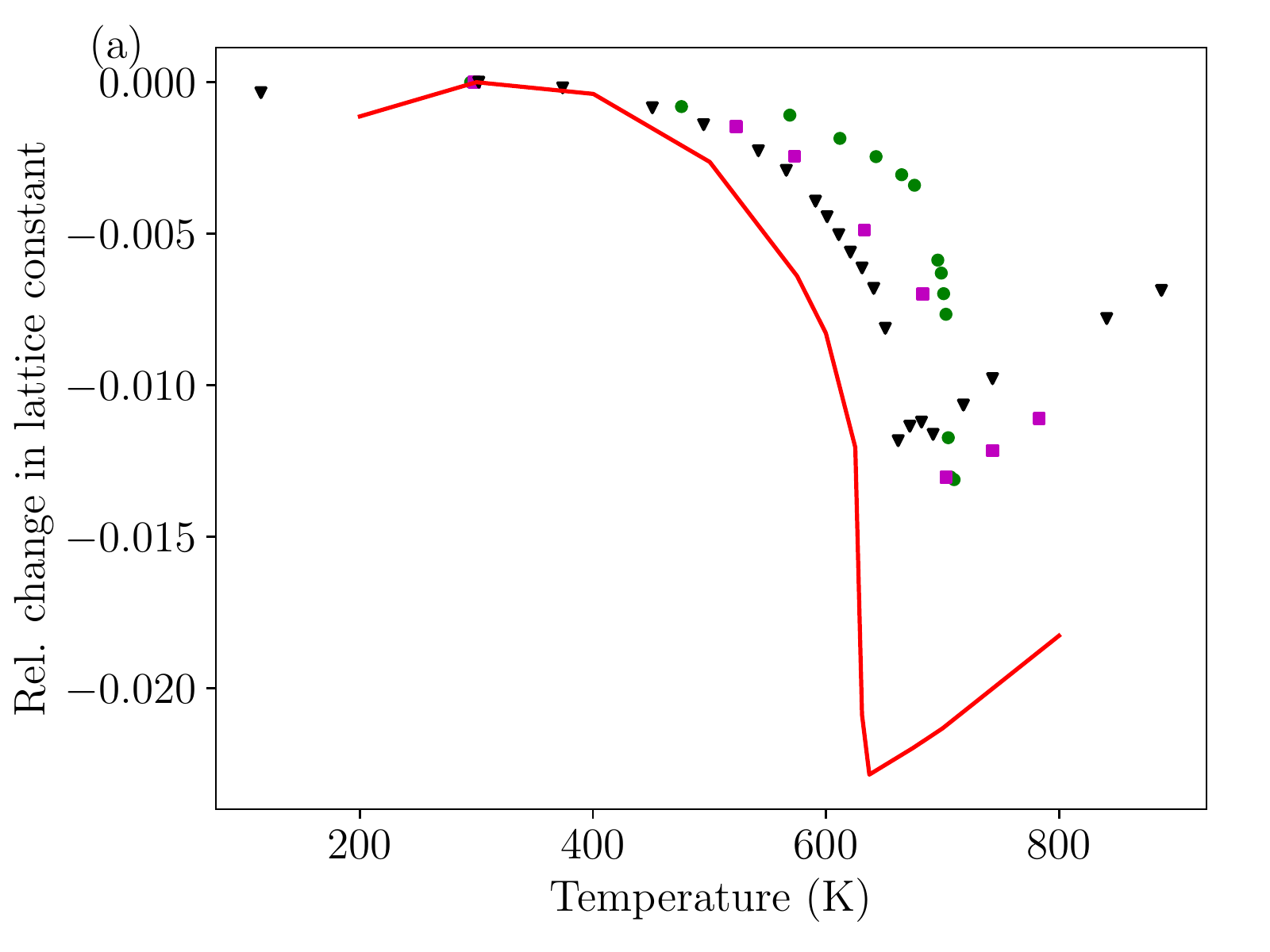}
\includegraphics[width=0.9\linewidth]{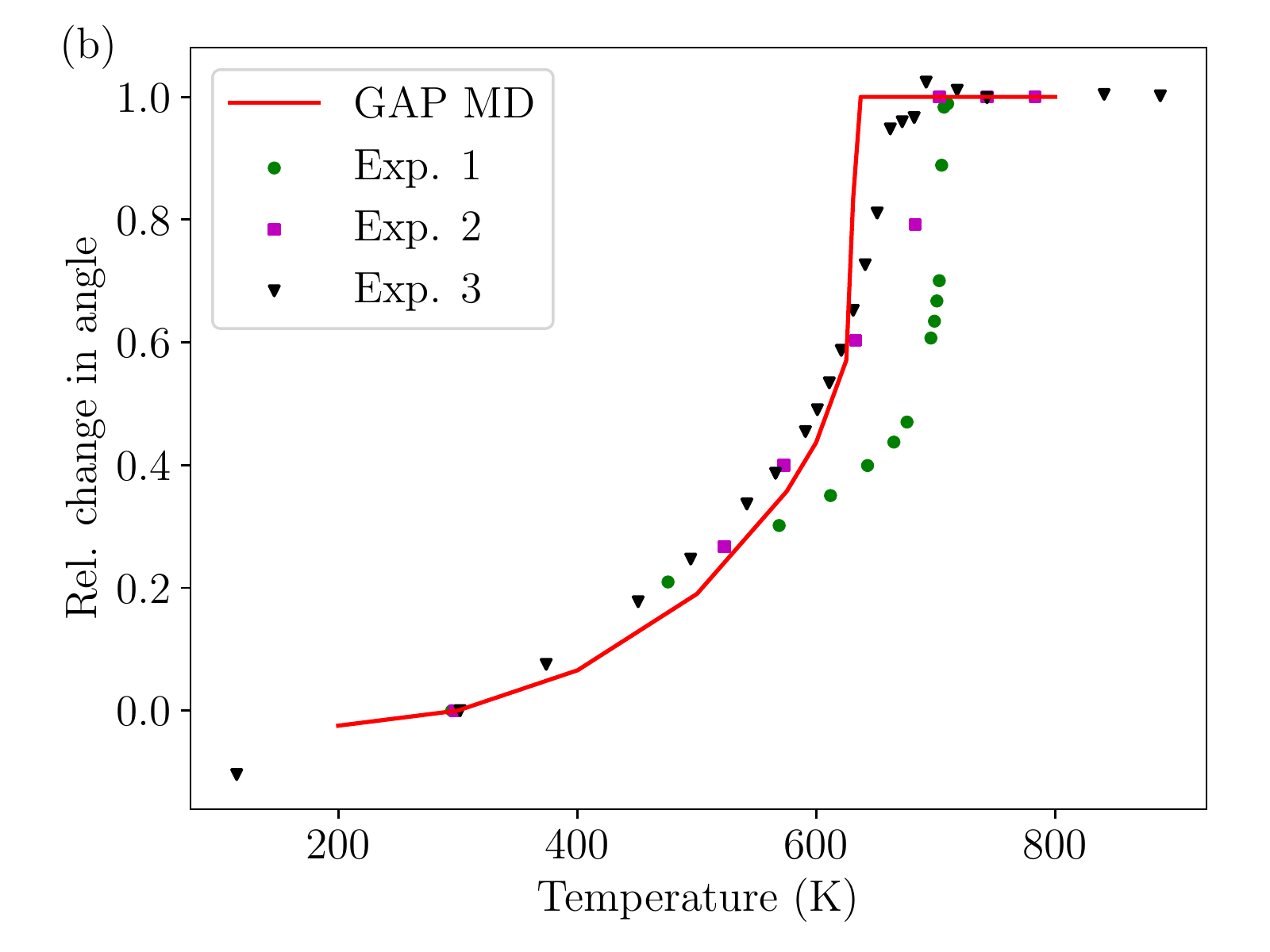}
\includegraphics[width=0.9\linewidth]{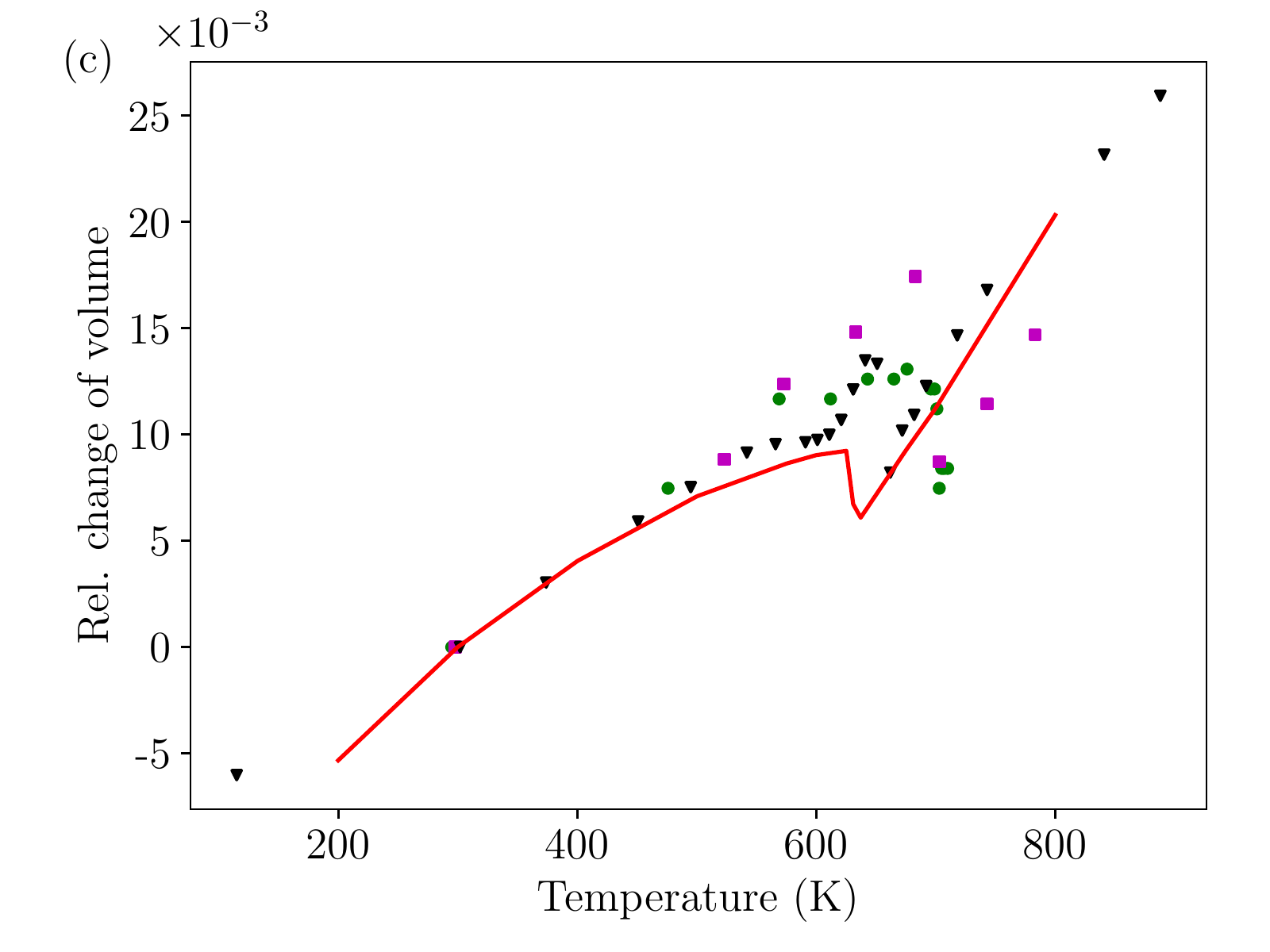}
\caption{Relative change of (a) the lattice constant, (b) the rhombohedral angle and (c) the volume of GeTe with temperature. The red lines are our MD results, while the points represent experimental data taken from Refs.~\cite{main} (green), \cite{mdpigete} (magenta) and \cite{GeTeBo} (black).}
\label{fig1}
\end{center}
\end{figure}

We calculate the structural parameters of GeTe (the lattice constant $a$ and the rhombohedral angle $\theta$) at several temperatures. At each time step we calculate the instantaneous values of the lattice constant and rhombohedral angle from the geometry and volume of the simulation region. Following that, we find the structural parameters at a given temperature as a simple arithmetic mean of the instantaneous values along the MD trajectory. The results are given in Figure \ref{fig1} and compared with a number of available experiments~\cite{main, mdpigete, GeTeBo}. In this figure, we show the relative change of the structural parameters compared to their 300 K values ($V$ is the volume of the primitive cell):
\begin{align*}
\alpha _u &= \frac{u_T - u_{300\text{K}}}{u_{300\text{K}}}, \quad \text{for}\quad u = a, V, \\ 
\alpha _{\theta} &= \frac{\theta _T - \theta _{300\text{K}}}{60^{\circ} - \theta _{300\text{K}}}. \numberthis
\label{structchange}
\end{align*}

Our calculations reproduce the experimental results very well. All studies show negative expansion of the lattice constant at intermediate temperatures (above 300 K and below the critical temperature) and positive expansion in the cubic phase. We find that the lattice constant has a positive thermal expansion coefficient for temperatures below 300 K, as measured in the experiment. The rhombohedral angle tends to the cubic value of 60$^{\circ}$ at high temperatures. From the behavior of the rhombohedral angle, we can infer that the critical temperature in our study is 634 K (the middle point between the last rhombohedral structure at 631 K and the first cubic structure at 637 K), which is in the range of experimental results (600 - 700 K)~\cite{TCnote}.

We also calculate the volumetric thermal expansion of GeTe (see Fig.~\ref{fig1} (c)). Again, our results follow closely experimental findings, both showing negative thermal expansion (NTE) at the phase transition. In the cubic phase, GeTe regains positive thermal expansion, in agreement with experiment. We do not see a discontinuity in the calculated thermal expansion of GeTe. To be precise, we see a decrease of the volumetric thermal expansion coefficient as we approach the phase transition from lower temperatures, which eventually becomes negative thermal expansion at 631 K. For more elaborate discussion of negative thermal expansion near the phase transition, please see Supplementary material ~\cite{supp}.

\begin{figure}
\begin{center}
\includegraphics[width=0.9\linewidth]{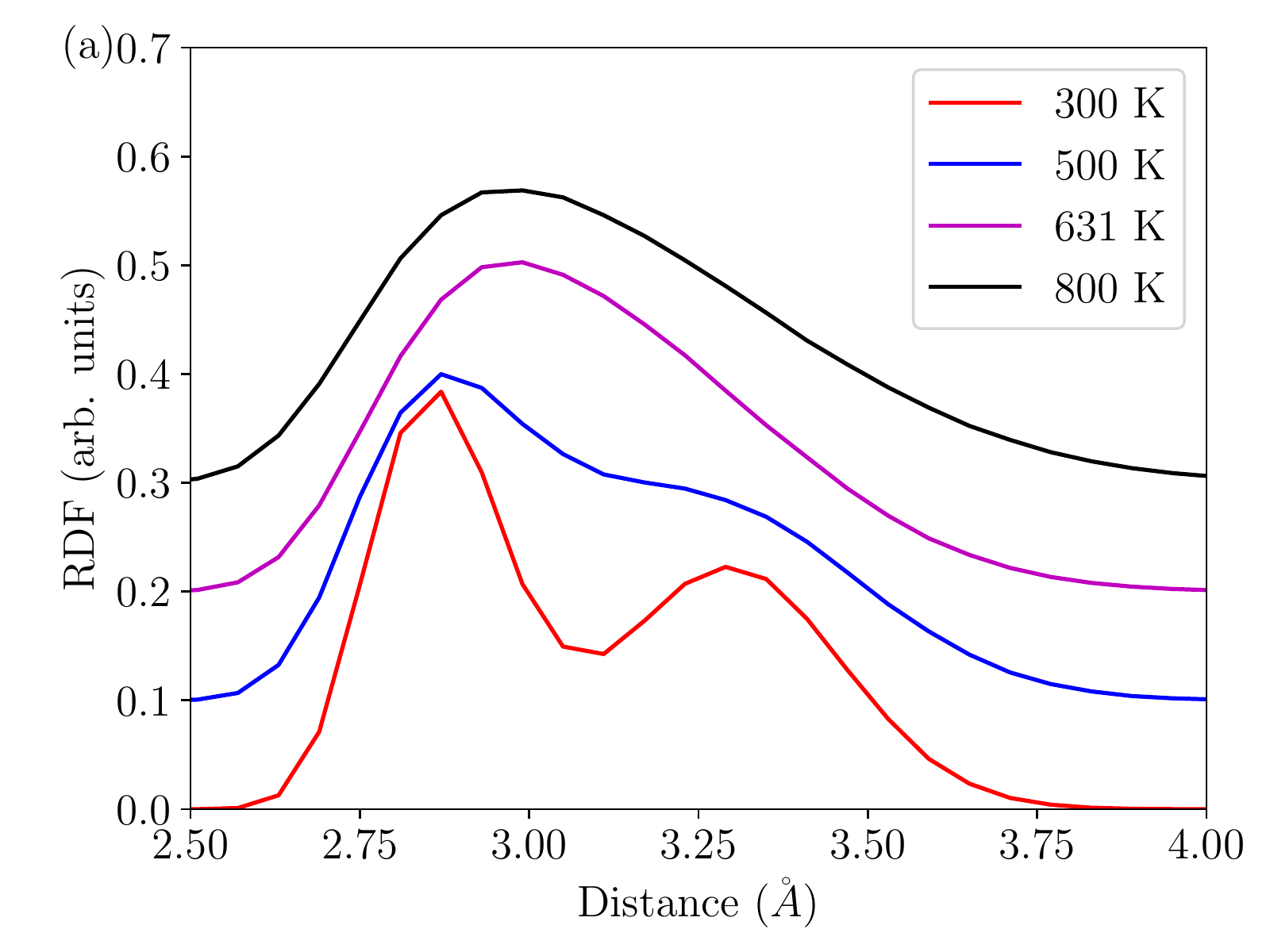}
\includegraphics[width=0.9\linewidth]{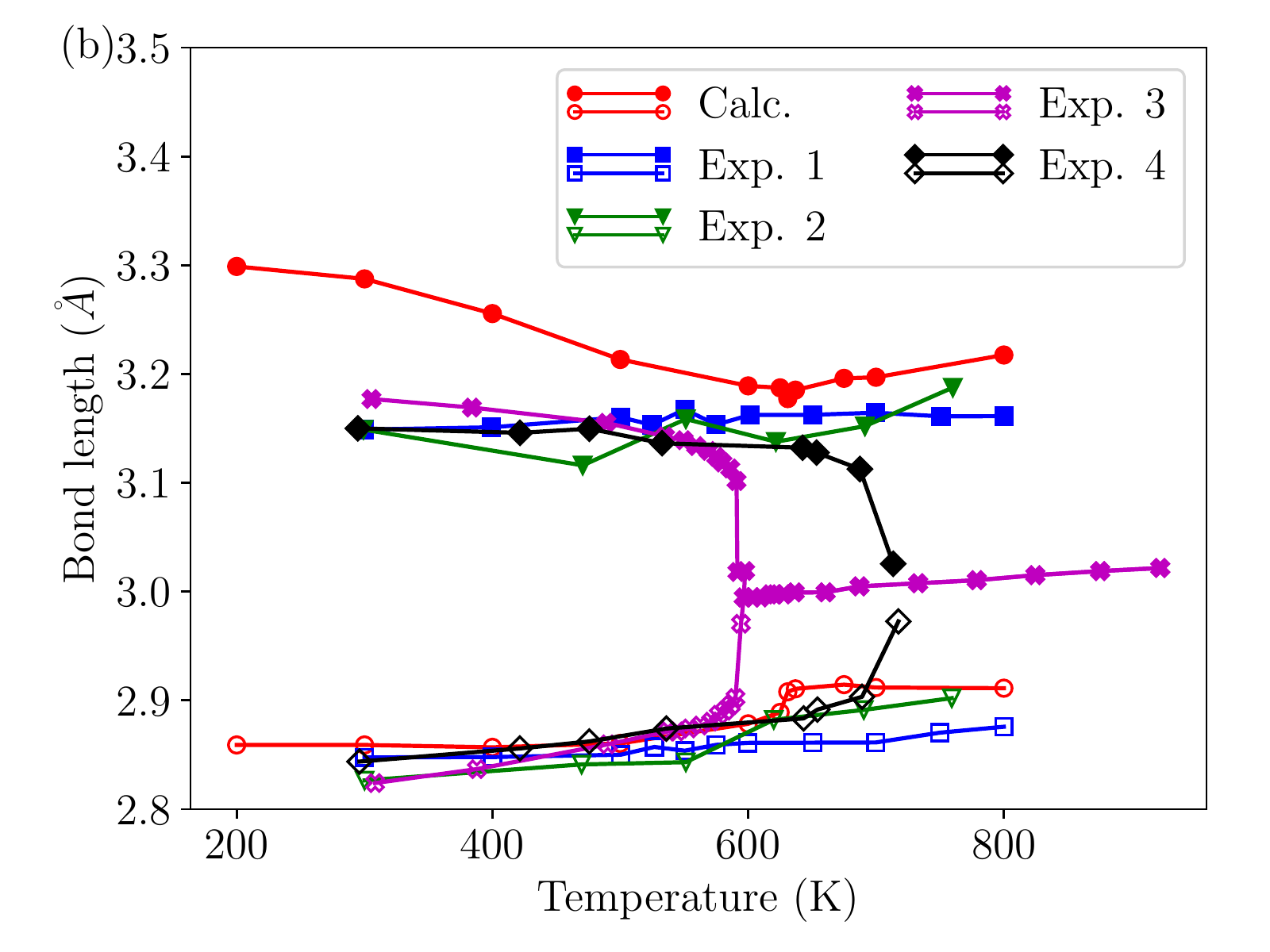}
\caption{(a) Radial distribution function of GeTe for the nearest-neighbor bonds at different temperatures. (b) Nearest-neighbor bond lengths in GeTe. Our calculations are in red, Ref.~\cite{Fons} is in blue, Ref.~\cite{macunaga} is in green points (experiments reporting the order-disorder character of the phase transition). Ref.~\cite{newmain} is in magenta and Ref.~\cite{main} is in black points (experiments reporting the displacive character of the phase transition). The full symbols represent longer bond lengths, while the empty symbols represent shorter bond lengths. The lines are guides to the eye.}
\label{fig3}
\end{center}
\end{figure}

Finally, we compute the Ge - Te nearest-neighbor bond lengths. A number of experimental~\cite{Fons, macunaga} and theoretical studies~\cite{ChatterjiMD, GeTeAIMD} claim that they observe persistence of unequal bond lengths in the cubic phase. Previous theoretical studies infer this effect from the distorted Gaussian shape of the radial distribution function (RDF) for these bond lengths ($\approx$ 3 \AA). If the interatomic interaction is perfectly harmonic, one would expect bond lengths to be normally distributed around some mean value which is the reported bond length. A distortion of this Gaussian shape is usually attributed to the presence of two Gaussians, which means that we have two different bond lengths in the considered length scale. Similarly to the previous theoretical studies, we also find two different bond lengths if we try to fit our data with two Gaussians (see Figure~\ref{fig3} (a)). Figure~\ref{fig3} (b) shows the fitted bond lengths using two Gaussians in our calculation compared to experiments~\citep{main, newmain, Fons, macunaga}. The experiments that can probe local structure obtain unequal bond lengths in the cubic phase ~\cite{Fons, macunaga} (Exp. 1 and Exp. 2 in Figure~\ref{fig3} (b)), while the experiments that see the average structure see the equivalent bond lengths~\cite{main, newmain} (Exp. 3 and Exp. 4 in Figure~\ref{fig3} (b)). Our results obtained by fitting the RDF with two Gaussians are in overall agreement with experimental results that see the local structure. However,  contrary to those experiments, we see an interesting behavior near the phase transition, a noticeable increase in the short bond length and a noticeable decrease of the larger bond length. This change of bond lengths does not make them equal however, and the local rhombohedral phase seems to persist in the cubic phase as well. 

From the analysis above, it is clear why the non-Gaussian shape of the nearest-neighbor bond length has been interpreted as a fingerprint of the order-disorder phase transition~\cite{GeTeAIMD}. However, we observe the same non-Gaussian behavior even in the case of other rocksalt compounds modeled using interatomic potentials, such as PbTe~\cite{JavierMD} and MgO~\cite{MgO_MD} (see Supplementary material~\cite{supp}). These two materials are undoubtedly rocksalt and still have the distorted Gaussian shape for the nearest-neighbor RDF. The deviation from the simple Gaussian shape of the RDF is stronger in PbTe compared to MgO, probably because PbTe is more anharmonic. The reasoning for this assumption is as follows. First, we assume that the bond length is determined solely due to pairwise interactions between atoms. Then the probability distribution of that bond length would be proportional to $\exp(-U(R)/k_{B}T)$, where $U(R)$ is the energy of that two atom system. If this energy is purely harmonic, we would have a Gaussian distribution of the bond length. However, in case that the bond has an anharmonic term, there will be a skewing of the distribution in one of the directions, which is what we observe in all three systems (MgO, PbTe and GeTe). Additionally, we find that the fitting procedure fails to correctly reproduce the "static" bond lengths in the rhombohedral phase of GeTe (see Supplementary material~\cite{supp}). Hence, our results show that the non-Gaussian shape of the RDF is not a proof of the order-disorder behavior in GeTe and we are more inclined to believe it is a consequence of the large anharmonicity of the Ge-Te nearest-neighbor bond. 

\section{Order parameter}

We calculated properties of the order parameter at various temperatures in order to understand the driving mechanism for the phase transition in GeTe. We calculate the local order parameter ($\tau _{i}(t)$ for the $i$-th unit cell inside the supercell) as:
\begin{align*}
\tau _{i}(t) = \vec{x}_{Te, i}(t) - \vec{x}_{Ge, i}(t) - 0.5\sum _{j} \vec{R}_{j}, \numberthis \label{orderparameter}
\end{align*}
where $\vec{x}_{Te/Ge, i}$ is the instantaneous position of the tellurium/germanium atom in the $i$th unit cell and $\vec{R}_{j}$ are the primitive lattice vectors at a given temperature. The average over all unit cells inside the MD simulation region at a certain time step represents the instantaneous order parameter. The average of the instantaneous order parameters over the entire MD trajectory represents the order paramater for that temperature (see Supplementary material~\cite{supp} for additional information).

\begin{figure}
\begin{center}
\includegraphics[width=0.9\linewidth]{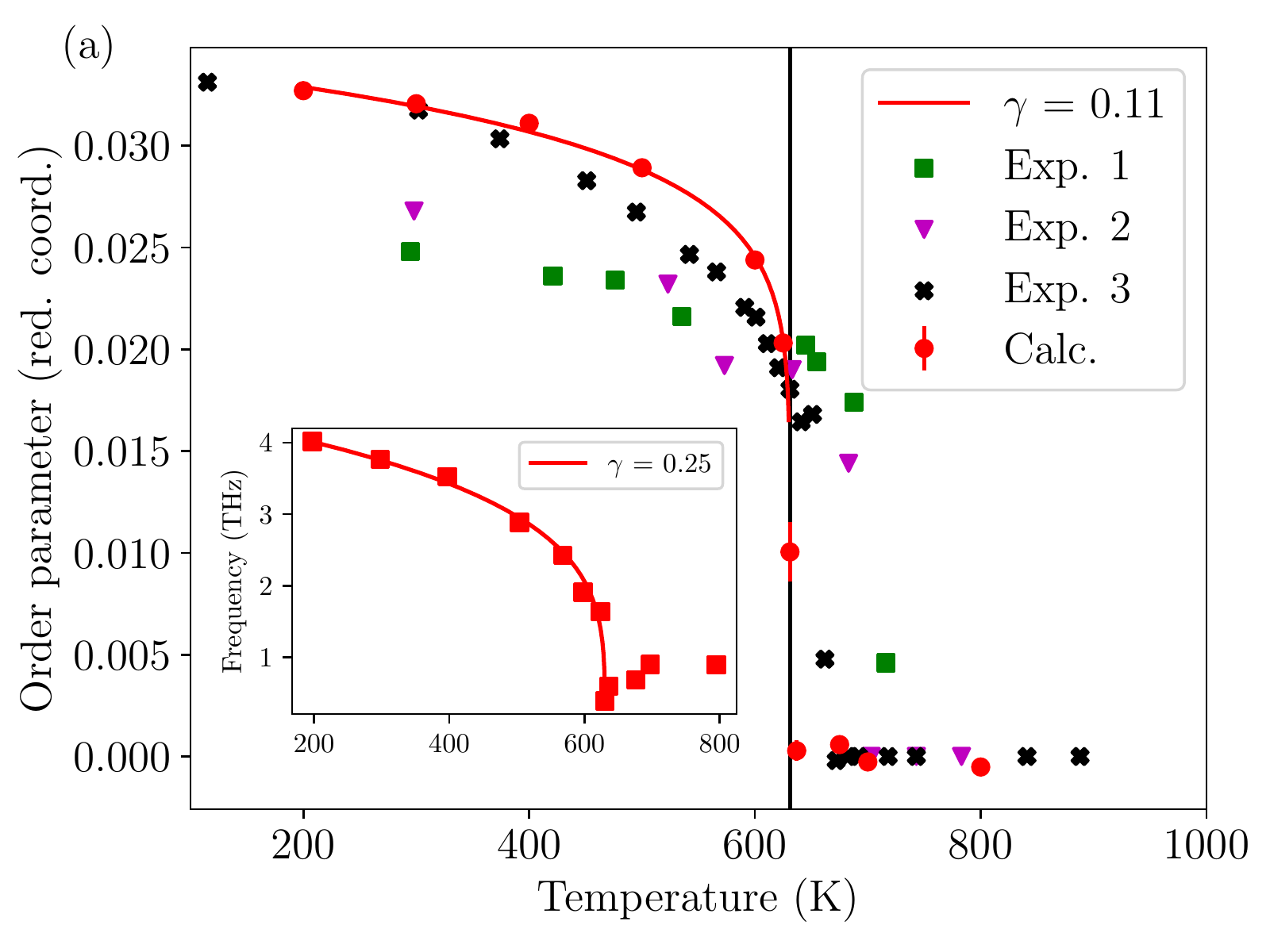}
\includegraphics[width=0.9\linewidth]{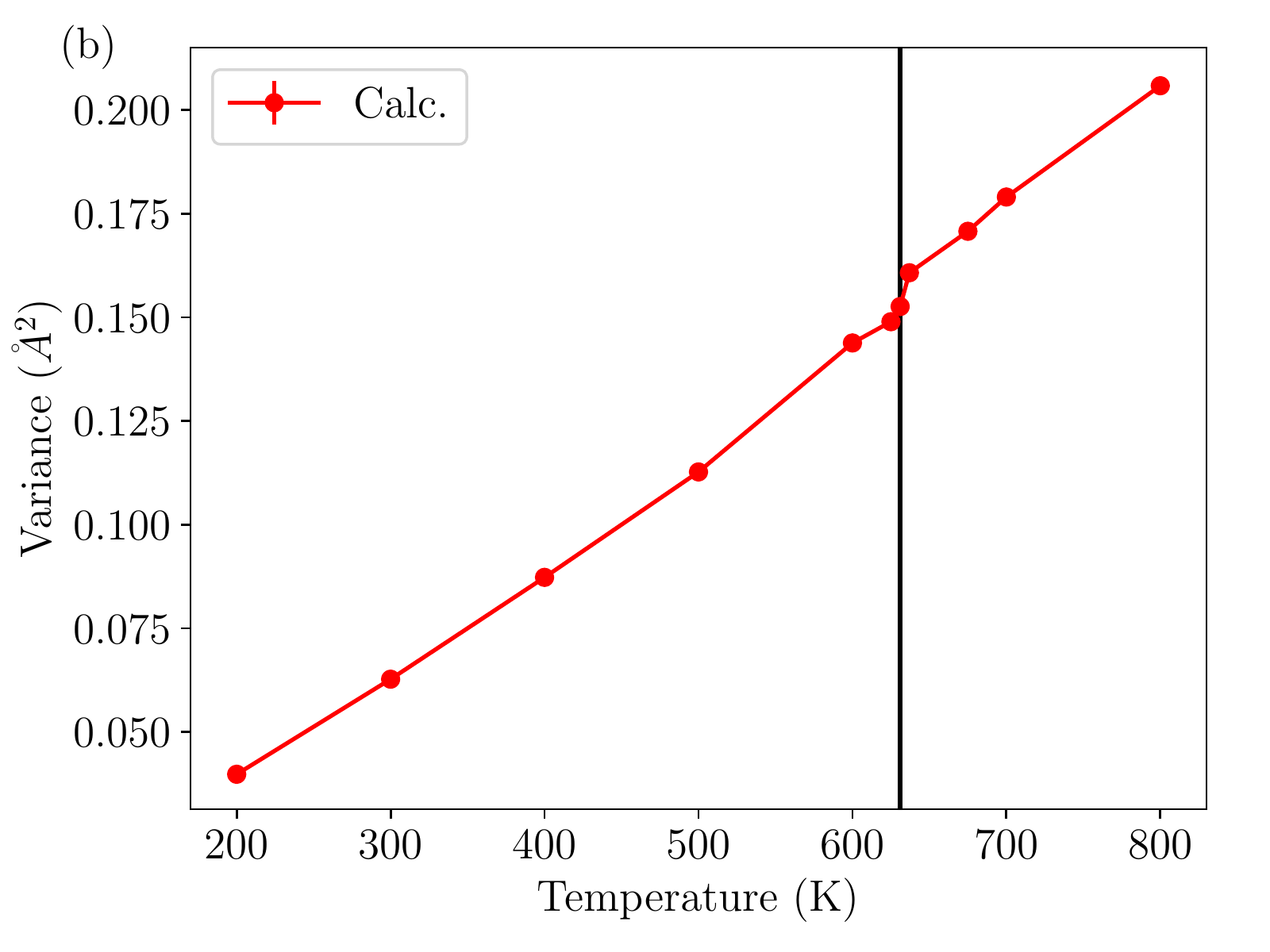}
\caption{Temperature dependence of (a) the average order parameter and (b) the variance of the order parameter as extracted from the probability distribution functions of the order parameter at different times and temperatures (see Supplementary material~\cite{supp} for more explanation). The experimental points are taken from Refs.~\cite{main, mdpigete, GeTeBo}. The red points denote the results from our calculations, while the bars are the standard error of the averaged quantity. The vertical black line shows the phase transition boundary. In (a), the red solid lines correspond to the power law fit to our calculated results, where $\gamma$ is the exponent. The inset in (a) shows the calculated soft transverse optical phonon frequency versus temperature.}
\label{fig4}
\end{center}
\end{figure}

Figure \ref{fig4} (a) shows the total order parameter calculated at different temperatures and compared with available experimental literature~\cite{main, mdpigete, GeTeBo}. We can see that the overall agreement is good and that the differences mostly come from different values of the critical temperature. The inset shows the temperature dependence of the soft TO (A$_{1\text{g}}$) phonon mode frequency. We have calculated the phonon frequencies using the temperature dependent effective potential method~\cite{TDEP1, TDEP2, TDEP3}. We can fit the temperature dependence of the order parameter and the soft TO mode to a simple functional dependence:
\begin{align*}
f(T) = A(T_{C} - T)^{\gamma},
\end{align*}
where $T_{C}$ is the critical temperature, while $A$ and $\gamma$ are the fitting parameters. A simple Landau approach to the displacive phase transitions predicts that the $\gamma$ parameter for the soft TO mode should have the same value as the $\gamma$ exponent for the order parameter~\cite{landau} (see Supplementary material for clarification). This is not what we find in our calculations: $\gamma$ for the order parameter is around two times smaller than $\gamma$ for the soft TO mode. The possible reason for this is that the effects of the other degrees of freedom besides the order parameter (such as large strain-order parameter coupling or disorder of the local order parameter) make a straightforward consideration of the Landau model inapplicable.

The procedure we used to calculate the average order parameter at a certain temperature allows us to calculate the time average variance of the local order parameter (see Supplementary material~\cite{supp} for clarification), which is shown in Fig.~\ref{fig4} (b). We can notice a small jump at the phase transition. The standard errors are shown in the figure and they are much smaller than the size of the step in variance at the phase transition (the standard errors are smaller than the points). This increase is similar to the increase observed for the Debye-Waller factor in experiments. In experiments, this increase is explained by the ambiguity in determining the crystallographic phase of the system, i.e. whether the system is in a unique phase or mixing the two phases~\cite{GeTeBo}. 

We suggest that the increase in the variance in Fig. \ref{fig4} (b) is due to a weak order-disorder character of the phase transition. The model for this behavior is as follows. The unit cell polarizations in our simulations are distributed according to a normal distribution centered around the instantaneous order parameter. The variance of that distribution should be a smooth function of temperature. In the displacive phase transition, the instantaneous order parameter becomes zero and there should be no abrupt change in the variance. However, in the order-disorder phase transition, the unit cell's polarization would be normally distributed around two values of the instantaneous order parameter that have the same absolute magnitude but opposite signs. If these two mean values are sufficiently close, the distributions of the unit cell order parameter would overlap and yield a single peak behavior, obscuring the order-disorder character. However, in this scenario the variance of the unit cell polarization distribution may abruptly change, which is what we calculate in our simulations. Hence, we interpret the observed step in the variance as the signature of a weak order-disorder character of the phase transition.

Finally, we look at the dynamics of the order parameter at different temperatures. We define the order parameter correlation function as:
\begin{align*}
&G_{\alpha\beta}(\vec{r},t) = \langle \tau _{\alpha} (0,0)\tau _{\beta}(\vec{r},t)\rangle = \\
&\int\int\text{d}\vec{r}'\text{d}t' \left(\tau _{\alpha} (\vec{r}',t') - \langle\tau _{\alpha}\rangle\right)\left(\tau _{\beta} (\vec{r}' + \vec{r},t' + t) - \langle\tau _{\beta}\rangle\right). \numberthis \label{opcf}
\end{align*}
Here $\alpha ,\beta$ denote the Cartesian coordinates and $\vec{r}$ is the position vector of the unit cell. We can find the Fourier transform of this quantity, $\Gamma _{\alpha\beta} (\vec{q}, \omega)$. 

\begin{figure}
\begin{center}
\includegraphics[width=0.9\linewidth]{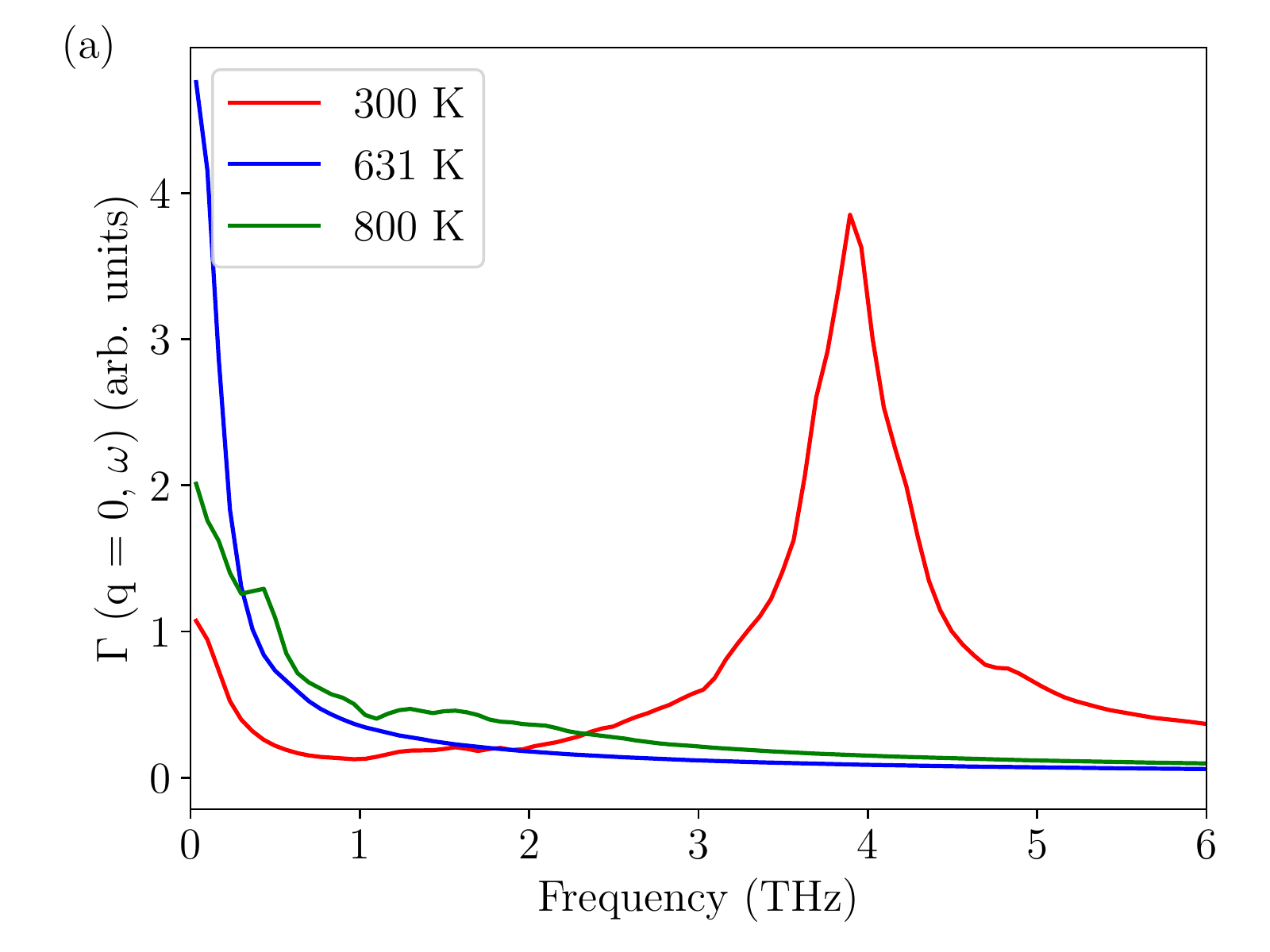}
\includegraphics[width=0.9\linewidth]{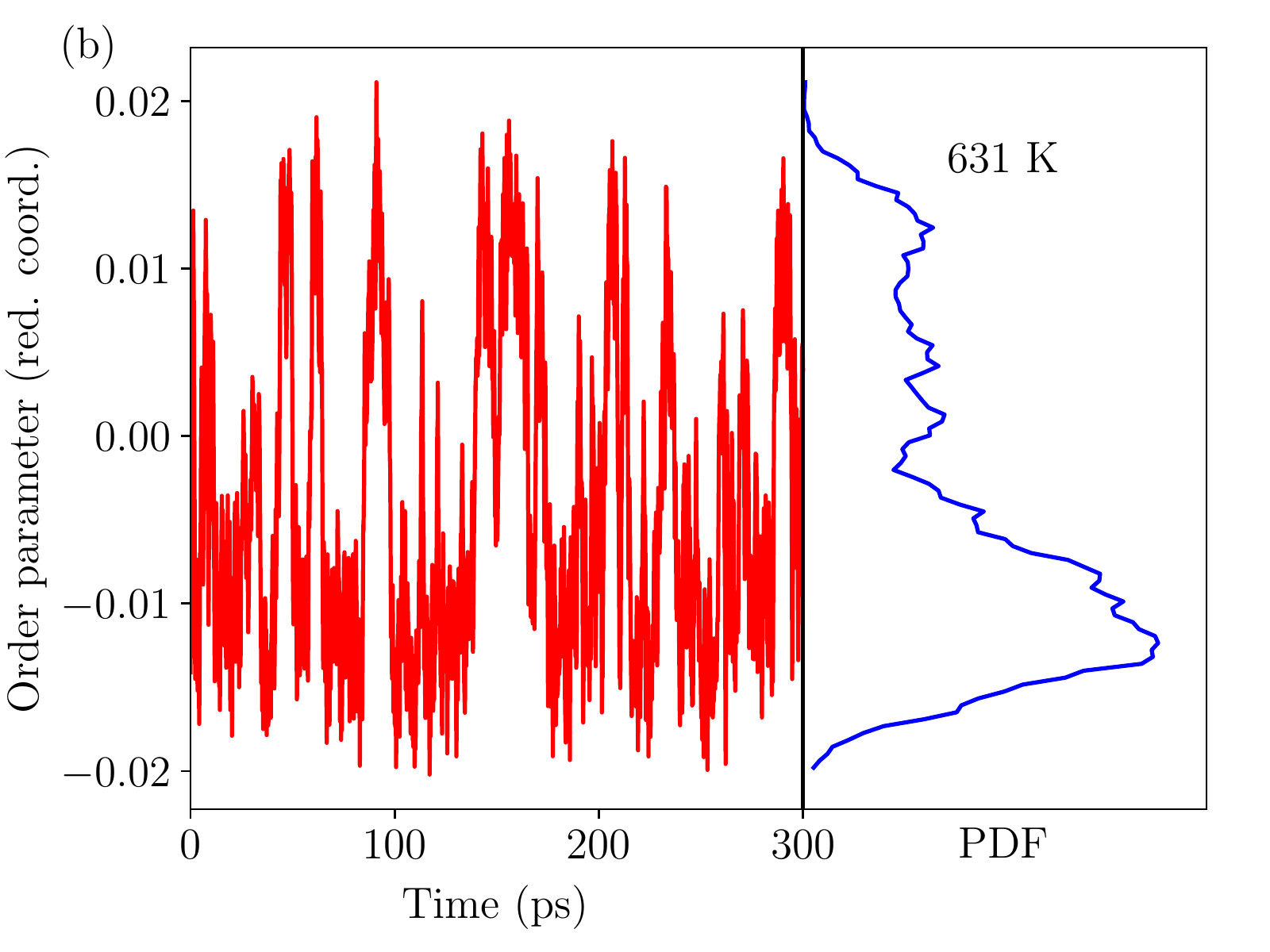}
\caption{(a) Frequency dependence of the Fourier transform of the order parameter correlation function (see Eq.~\ref{opcf}) at different temperatures. (b) The time evolution of the order parameter at 631 K. The side plot shows the probability density function of the instantaneous order parameter at this temperature.}
\label{fig5}
\end{center}
\end{figure}

First, we discuss the behavior of $\Gamma _{zz}(\vec{q} = 0, \omega)$ at different temperatures, shown in Figure~\ref{fig5} (a). We oriented our simulation cell so that the order parameter is polarized along the $z$ Cartesian direction. We can see that at low temperatures (300 K), the peak of this quantity is around 4 THz, which is the frequency of the soft optical mode, see Fig.~\ref{fig4} (a). Additionally, we find that the order parameter correlation function in the other two Cartesian directions has a peak at the frequency of other optical modes (see Supplementary material~\cite{supp}). This peak softens as we approach the transition temperature and disappears at the phase transition, leaving only a quasielastic peak (i.e. peak at the zero frequency, corresponding to zero energy transfer in scattering experiments). Interestingly, in the cubic phase, only the quasielastic peak persists and the oscillations of the order parameter cannot be associated with any phonon. This behavior shows there is no persistent correlation among unit cell polarizations throughout the simulation region, suggesting a displacive character of the phase transition. On the other hand, in the case of the order-disorder phase transition, the thermal oscilations of the polarization around local minima, although in opposite wells and out of phase, would still oscillate with the same frequency, leading to a non-zero signal in Fig.~\ref{fig4} (a).

To understand better the dynamics of the order parameter at the phase transition, we show the instantaneous order parameter $\langle \tau _{z} \rangle (t)$ at 631 K in Figure \ref{fig5} (b). We can see that although the spatial correlation persists and the order parameter is non-zero (the system is still in the rhombohedral phase), the value of the order parameter starts to switch between plus and minus signs. This switching leads to an exponentially decaying correlation function and ultimately to the quasielastic peak observed in Fig. \ref{fig5} (a). The switching behavior is dependent on the simulation cell size, with larger cells having lower frequencies of switching. 
This indicates that in the thermodynamic limit there would be no switching and thus this behavior cannot be interpreted as the order-disorder phase transition.

We also calculated the order parameter correlation length at different temperatures. We observed a large jump of the correlation length at the phase transition (see Supplementary material~\cite{supp}), which suggests that the phase transition is of the second order. However, due to the small size of our simulation cell, we could not observe whether the divergence of the correlation length follows a specific power law dependence. 

\section{Conclusions}

In summary, we used molecular dynamics simulations to model the structural phase transition in GeTe using our recently developed Gaussian Approximation Potential for GeTe that mimics density functional theory results very well. First, we confirmed the results of our recent study on negative thermal expansion at the phase transition. 

Secondly, we calculated the radial distribution function of the nearest-neighbor bonds in GeTe for both crystallographic phases. We observed a strongly non-Gaussian shape of the radial distribution function in the cubic phase in accordance to previous ab-initio MD simulations. However, we showed that this effect is most probably due to the strong anharmonicity of the Ge-Te bonds and not due to a persistent local rhombohedral distortion in the cubic phase. 

Next, we discussed in detail the behavior of the order parameter at the phase transition. We found fingerprints of both order-disorder and displacive phase transitions. Both the order parameter and the soft TO mode frequency continuously fall to zero at the phase transition, pointing to the displacive character of the phase transition. However, the variance of the order parameter probability distribution function exhibits a small step at the phase transition, which can be explained by the order-disorder phase transition. 

Finally, we investigated the dynamics of the order parameter at the phase transition and found that in the low symmetry phase it closely follows the behavior of the soft TO mode. This correlation disappears in the cubic phase, suggesting a possible displacive character of the phase transition. We showed the emergence of switching behavior at the phase transition, where polarization retains spatial correlation but loses temporal correlation. In the end, we calculated the order parameter correlation length for different temperatures and found that it diverges at the phase transition, suggesting that the phase transition is of the second order.

The distinction between the displacive and order-disorder phase transitions comes from the limiting cases of the simple Landau model of the second-order phase transition. As simple ground state calculations show~\citep{BrucePT}, GeTe does not belong to either of these limiting cases, although it can be shown to be closer to the displacive model. The lengthy and detailed investigation that we carried out using molecular dynamics confirms this conclusion. The displacive character of the phase transition is supported by the disappearance of the temporal correlation of the order parameter with the persisting spatial correlation. A weak order-disorder character of the phase transition can be inferred from the temperature dependence of the variance of the order parameter.  

Germanium telluride is a rare material where both the order-disorder and the displacive character of the phase transition coexist. Similar behavior might exist in lead chalcogenides, where the off-centering of local dipoles has been claimed to be observed~\cite{PbTeMD, PbTe1, PbTe2}. This effect could occur also in ferroelectric SnTe~\cite{SnTe1}. A detailed first principle study has disproved the existence of off-centering in PbTe~\cite{PbTeMD}. Clearly, IV-VI materials represent very interesting test subjects for this type of study. While there has been a fairly large number of papers on the properties of PbTe and GeTe, the other members of this group have not received as much attention. Finally, it would be interesting to see whether the coexistence of order-disorder and displacive character of the phase transition has an influence on the phonon and transport properties of GeTe, specifically lattice thermal conductivity. While our previous study~\cite{OurTC} resolved the enigma of the lattice thermal conductivity enhancement at the phase transition, it did not capture the influence of the order-disorder character.

\section{Acknowledgements}

This work is supported by Science Foundation Ireland under grant numbers 15/IA/3160 and 13/RC/2077. The later grant is co-funded under the European Regional Development Fund. We acknowledge the Irish Centre for High-End Computing (ICHEC) for the provision of computational facilities.


\begin{thebibliography}{46}%
\makeatletter
\providecommand \@ifxundefined [1]{%
 \@ifx{#1\undefined}
}%
\providecommand \@ifnum [1]{%
 \ifnum #1\expandafter \@firstoftwo
 \else \expandafter \@secondoftwo
 \fi
}%
\providecommand \@ifx [1]{%
 \ifx #1\expandafter \@firstoftwo
 \else \expandafter \@secondoftwo
 \fi
}%
\providecommand \natexlab [1]{#1}%
\providecommand \enquote  [1]{``#1''}%
\providecommand \bibnamefont  [1]{#1}%
\providecommand \bibfnamefont [1]{#1}%
\providecommand \citenamefont [1]{#1}%
\providecommand \href@noop [0]{\@secondoftwo}%
\providecommand \href [0]{\begingroup \@sanitize@url \@href}%
\providecommand \@href[1]{\@@startlink{#1}\@@href}%
\providecommand \@@href[1]{\endgroup#1\@@endlink}%
\providecommand \@sanitize@url [0]{\catcode `\\12\catcode `\$12\catcode
  `\&12\catcode `\#12\catcode `\^12\catcode `\_12\catcode `\%12\relax}%
\providecommand \@@startlink[1]{}%
\providecommand \@@endlink[0]{}%
\providecommand \url  [0]{\begingroup\@sanitize@url \@url }%
\providecommand \@url [1]{\endgroup\@href {#1}{\urlprefix }}%
\providecommand \urlprefix  [0]{URL }%
\providecommand \Eprint [0]{\href }%
\providecommand \doibase [0]{https://doi.org/}%
\providecommand \selectlanguage [0]{\@gobble}%
\providecommand \bibinfo  [0]{\@secondoftwo}%
\providecommand \bibfield  [0]{\@secondoftwo}%
\providecommand \translation [1]{[#1]}%
\providecommand \BibitemOpen [0]{}%
\providecommand \bibitemStop [0]{}%
\providecommand \bibitemNoStop [0]{.\EOS\space}%
\providecommand \EOS [0]{\spacefactor3000\relax}%
\providecommand \BibitemShut  [1]{\csname bibitem#1\endcsname}%
\let\auto@bib@innerbib\@empty
\bibitem [{\citenamefont {Steigmeier}\ and\ \citenamefont
  {Harbeke}(1970)}]{STEIGMEIER}%
  \BibitemOpen
  \bibfield  {author} {\bibinfo {author} {\bibfnamefont {E.}~\bibnamefont
  {Steigmeier}}\ and\ \bibinfo {author} {\bibfnamefont {G.}~\bibnamefont
  {Harbeke}},\ }\bibfield  {title} {\bibinfo {title} {Soft phonon mode and
  ferroelectricity in {GeTe}},\ }\href
  {https://doi.org/https://doi.org/10.1016/0038-1098(70)90619-8} {\bibfield
  {journal} {\bibinfo  {journal} {Solid State Commun.}\ }\textbf {\bibinfo
  {volume} {8}},\ \bibinfo {pages} {1275 } (\bibinfo {year}
  {1970})}\BibitemShut {NoStop}%
\bibitem [{\citenamefont {Chattopadhyay}\ \emph {et~al.}(1987)\citenamefont
  {Chattopadhyay}, \citenamefont {Boucherle},\ and\ \citenamefont
  {vonSchnering}}]{main}%
  \BibitemOpen
  \bibfield  {author} {\bibinfo {author} {\bibfnamefont {T.}~\bibnamefont
  {Chattopadhyay}}, \bibinfo {author} {\bibfnamefont {J.~X.}\ \bibnamefont
  {Boucherle}},\ and\ \bibinfo {author} {\bibfnamefont {H.~G.}\ \bibnamefont
  {vonSchnering}},\ }\bibfield  {title} {\bibinfo {title} {Neutron diffraction
  study on the structural phase transition in {GeTe}},\ }\href
  {http://stacks.iop.org/0022-3719/20/i=10/a=012} {\bibfield  {journal}
  {\bibinfo  {journal} {J. Phys. C: Solid State Phys.}\ }\textbf {\bibinfo
  {volume} {20}},\ \bibinfo {pages} {1431} (\bibinfo {year}
  {1987})}\BibitemShut {NoStop}%
\bibitem [{\citenamefont {Chatterji}\ \emph {et~al.}(2015)\citenamefont
  {Chatterji}, \citenamefont {Kumar},\ and\ \citenamefont {Wdowik}}]{newmain}%
  \BibitemOpen
  \bibfield  {author} {\bibinfo {author} {\bibfnamefont {T.}~\bibnamefont
  {Chatterji}}, \bibinfo {author} {\bibfnamefont {C.~M.~N.}\ \bibnamefont
  {Kumar}},\ and\ \bibinfo {author} {\bibfnamefont {U.~D.}\ \bibnamefont
  {Wdowik}},\ }\bibfield  {title} {\bibinfo {title} {Anomalous
  temperature-induced volume contraction in {GeTe}},\ }\href
  {https://doi.org/10.1103/PhysRevB.91.054110} {\bibfield  {journal} {\bibinfo
  {journal} {Phys. Rev. B}\ }\textbf {\bibinfo {volume} {91}},\ \bibinfo
  {pages} {054110} (\bibinfo {year} {2015})}\BibitemShut {NoStop}%
\bibitem [{\citenamefont {Wdowik}\ \emph {et~al.}(2014)\citenamefont {Wdowik},
  \citenamefont {Parlinski}, \citenamefont {Rols},\ and\ \citenamefont
  {Chatterji}}]{displejsiv}%
  \BibitemOpen
  \bibfield  {author} {\bibinfo {author} {\bibfnamefont {U.~D.}\ \bibnamefont
  {Wdowik}}, \bibinfo {author} {\bibfnamefont {K.}~\bibnamefont {Parlinski}},
  \bibinfo {author} {\bibfnamefont {S.}~\bibnamefont {Rols}},\ and\ \bibinfo
  {author} {\bibfnamefont {T.}~\bibnamefont {Chatterji}},\ }\bibfield  {title}
  {\bibinfo {title} {Soft-phonon mediated structural phase transition in
  {GeTe}},\ }\href {https://doi.org/10.1103/PhysRevB.89.224306} {\bibfield
  {journal} {\bibinfo  {journal} {Phys. Rev. B}\ }\textbf {\bibinfo {volume}
  {89}},\ \bibinfo {pages} {224306} (\bibinfo {year} {2014})}\BibitemShut
  {NoStop}%
\bibitem [{\citenamefont {Fons}\ \emph {et~al.}(2010)\citenamefont {Fons},
  \citenamefont {Kolobov}, \citenamefont {Krbal}, \citenamefont {Tominaga},
  \citenamefont {Andrikopoulos}, \citenamefont {Yannopoulos}, \citenamefont
  {Voyiatzis},\ and\ \citenamefont {Uruga}}]{Fons}%
  \BibitemOpen
  \bibfield  {author} {\bibinfo {author} {\bibfnamefont {P.}~\bibnamefont
  {Fons}}, \bibinfo {author} {\bibfnamefont {A.~V.}\ \bibnamefont {Kolobov}},
  \bibinfo {author} {\bibfnamefont {M.}~\bibnamefont {Krbal}}, \bibinfo
  {author} {\bibfnamefont {J.}~\bibnamefont {Tominaga}}, \bibinfo {author}
  {\bibfnamefont {K.~S.}\ \bibnamefont {Andrikopoulos}}, \bibinfo {author}
  {\bibfnamefont {S.~N.}\ \bibnamefont {Yannopoulos}}, \bibinfo {author}
  {\bibfnamefont {G.~A.}\ \bibnamefont {Voyiatzis}},\ and\ \bibinfo {author}
  {\bibfnamefont {T.}~\bibnamefont {Uruga}},\ }\bibfield  {title} {\bibinfo
  {title} {Phase transition in crystalline {GeTe}: Pitfalls of averaging
  effects},\ }\href {https://doi.org/10.1103/PhysRevB.82.155209} {\bibfield
  {journal} {\bibinfo  {journal} {Phys. Rev. B}\ }\textbf {\bibinfo {volume}
  {82}},\ \bibinfo {pages} {155209} (\bibinfo {year} {2010})}\BibitemShut
  {NoStop}%
\bibitem [{\citenamefont {Matsunaga}\ \emph {et~al.}(2011)\citenamefont
  {Matsunaga}, \citenamefont {Fons}, \citenamefont {Kolobov}, \citenamefont
  {Tominaga},\ and\ \citenamefont {Yamada}}]{macunaga}%
  \BibitemOpen
  \bibfield  {author} {\bibinfo {author} {\bibfnamefont {T.}~\bibnamefont
  {Matsunaga}}, \bibinfo {author} {\bibfnamefont {P.}~\bibnamefont {Fons}},
  \bibinfo {author} {\bibfnamefont {A.~V.}\ \bibnamefont {Kolobov}}, \bibinfo
  {author} {\bibfnamefont {J.}~\bibnamefont {Tominaga}},\ and\ \bibinfo
  {author} {\bibfnamefont {N.}~\bibnamefont {Yamada}},\ }\bibfield  {title}
  {\bibinfo {title} {The order-disorder transition in {GeTe}: Views from
  different length-scales},\ }\href {https://doi.org/10.1063/1.3665067}
  {\bibfield  {journal} {\bibinfo  {journal} {Appl. Phys. Lett.}\ }\textbf
  {\bibinfo {volume} {99}},\ \bibinfo {pages} {231907} (\bibinfo {year}
  {2011})}\BibitemShut {NoStop}%
\bibitem [{\citenamefont {Landau}\ and\ \citenamefont
  {Lifshitz}(2013)}]{landau}%
  \BibitemOpen
  \bibfield  {author} {\bibinfo {author} {\bibfnamefont {L.}~\bibnamefont
  {Landau}}\ and\ \bibinfo {author} {\bibfnamefont {E.}~\bibnamefont
  {Lifshitz}},\ }\href {https://books.google.ie/books?id=VzgJN-XPTRsC} {\emph
  {\bibinfo {title} {Statistical Physics}}},\ \bibinfo {number} {v. 5}\
  (\bibinfo  {publisher} {Elsevier Science},\ \bibinfo {year}
  {2013})\BibitemShut {NoStop}%
\bibitem [{TCn()}]{TCnote}%
  \BibitemOpen
  \href@noop {} {}\bibinfo {note} {The exact value of the transition
  temperature depends on the free charge carriers' concentration.}\BibitemShut
  {Stop}%
\bibitem [{\citenamefont {Levin}\ \emph {et~al.}(2013)\citenamefont {Levin},
  \citenamefont {Besser},\ and\ \citenamefont {Hanus}}]{levin}%
  \BibitemOpen
  \bibfield  {author} {\bibinfo {author} {\bibfnamefont {E.~M.}\ \bibnamefont
  {Levin}}, \bibinfo {author} {\bibfnamefont {M.~F.}\ \bibnamefont {Besser}},\
  and\ \bibinfo {author} {\bibfnamefont {R.}~\bibnamefont {Hanus}},\ }\bibfield
   {title} {\bibinfo {title} {Electronic and thermal transport in {GeTe}: A
  versatile base for thermoelectric materials},\ }\href@noop {} {\bibfield
  {journal} {\bibinfo  {journal} {J. Appl. Phys.}\ }\textbf {\bibinfo {volume}
  {114}},\ \bibinfo {pages} {083713} (\bibinfo {year} {2013})}\BibitemShut
  {NoStop}%
\bibitem [{\citenamefont {Wu}\ \emph {et~al.}(2017)\citenamefont {Wu},
  \citenamefont {Li}, \citenamefont {Wang}, \citenamefont {Zhang},
  \citenamefont {Yang}, \citenamefont {Zhang}, \citenamefont {Chen},\ and\
  \citenamefont {Yang}}]{Wu2017}%
  \BibitemOpen
  \bibfield  {author} {\bibinfo {author} {\bibfnamefont {L.}~\bibnamefont
  {Wu}}, \bibinfo {author} {\bibfnamefont {X.}~\bibnamefont {Li}}, \bibinfo
  {author} {\bibfnamefont {S.}~\bibnamefont {Wang}}, \bibinfo {author}
  {\bibfnamefont {T.}~\bibnamefont {Zhang}}, \bibinfo {author} {\bibfnamefont
  {J.}~\bibnamefont {Yang}}, \bibinfo {author} {\bibfnamefont {W.}~\bibnamefont
  {Zhang}}, \bibinfo {author} {\bibfnamefont {L.}~\bibnamefont {Chen}},\ and\
  \bibinfo {author} {\bibfnamefont {J.}~\bibnamefont {Yang}},\ }\bibfield
  {title} {\bibinfo {title} {Resonant level-induced high thermoelectric
  response in indium-doped {GeTe}},\ }\href@noop {} {\bibfield  {journal}
  {\bibinfo  {journal} {NPG Asia Mater.}\ }\textbf {\bibinfo {volume} {9}},\
  \bibinfo {pages} {e343} (\bibinfo {year} {2017})}\BibitemShut {NoStop}%
\bibitem [{\citenamefont {Li}\ \emph {et~al.}(2018)\citenamefont {Li},
  \citenamefont {Zhang}, \citenamefont {Chen}, \citenamefont {Lin},
  \citenamefont {Li}, \citenamefont {Shen}, \citenamefont {Witting},
  \citenamefont {Faghaninia}, \citenamefont {Chen}, \citenamefont {Jain},
  \citenamefont {Chen}, \citenamefont {Snyder},\ and\ \citenamefont
  {Pei}}]{pei-joule-gete}%
  \BibitemOpen
  \bibfield  {author} {\bibinfo {author} {\bibfnamefont {J.}~\bibnamefont
  {Li}}, \bibinfo {author} {\bibfnamefont {X.}~\bibnamefont {Zhang}}, \bibinfo
  {author} {\bibfnamefont {Z.}~\bibnamefont {Chen}}, \bibinfo {author}
  {\bibfnamefont {S.}~\bibnamefont {Lin}}, \bibinfo {author} {\bibfnamefont
  {W.}~\bibnamefont {Li}}, \bibinfo {author} {\bibfnamefont {J.}~\bibnamefont
  {Shen}}, \bibinfo {author} {\bibfnamefont {I.~T.}\ \bibnamefont {Witting}},
  \bibinfo {author} {\bibfnamefont {A.}~\bibnamefont {Faghaninia}}, \bibinfo
  {author} {\bibfnamefont {Y.}~\bibnamefont {Chen}}, \bibinfo {author}
  {\bibfnamefont {A.}~\bibnamefont {Jain}}, \bibinfo {author} {\bibfnamefont
  {L.}~\bibnamefont {Chen}}, \bibinfo {author} {\bibfnamefont {G.~J.}\
  \bibnamefont {Snyder}},\ and\ \bibinfo {author} {\bibfnamefont
  {Y.}~\bibnamefont {Pei}},\ }\bibfield  {title} {\bibinfo {title}
  {Low-symmetry rhombohedral {GeTe} thermoelectrics},\ }\href
  {http://www.sciencedirect.com/science/article/pii/S2542435118300850}
  {\bibfield  {journal} {\bibinfo  {journal} {Joule}\ } {\bibinfo  {volume} {2} }{\bibinfo  {issue} {5}\ } (\bibinfo {year}
  {2018})}\BibitemShut {NoStop}%
\bibitem [{\citenamefont {Perumal}\ \emph {et~al.}(2016)\citenamefont
  {Perumal}, \citenamefont {Roychowdhury},\ and\ \citenamefont
  {Biswas}}]{biswas-gete-rev}%
  \BibitemOpen
  \bibfield  {author} {\bibinfo {author} {\bibfnamefont {S.}~\bibnamefont
  {Perumal}}, \bibinfo {author} {\bibfnamefont {S.}~\bibnamefont
  {Roychowdhury}},\ and\ \bibinfo {author} {\bibfnamefont {K.}~\bibnamefont
  {Biswas}},\ }\bibfield  {title} {\bibinfo {title} {High performance
  thermoelectric materials and devices based on {GeTe}},\ }\href@noop {}
  {\bibfield  {journal} {\bibinfo  {journal} {J. Mater. Chem. C}\ }\textbf
  {\bibinfo {volume} {4}},\ \bibinfo {pages} {7520} (\bibinfo {year}
  {2016})}\BibitemShut {NoStop}%
\bibitem [{\citenamefont {Bruce}(1980)}]{BrucePT}%
  \BibitemOpen
  \bibfield  {author} {\bibinfo {author} {\bibfnamefont {A.~D.}\ \bibnamefont
  {Bruce}},\ }\bibfield  {title} {\bibinfo {title} {Structural phase
  transitions. {II}. static critical behaviour},\ }\href
  {https://doi.org/10.1080/00018738000101356} {\bibfield  {journal} {\bibinfo
  {journal} {Adv. Phys.}\ }\textbf {\bibinfo {volume} {29}},\ \bibinfo {pages}
  {111} (\bibinfo {year} {1980})}\BibitemShut {NoStop}%
\bibitem [{\citenamefont {Dangi\ifmmode~\acute{c}\else \'{c}\fi{}}\ \emph
  {et~al.}(2018)\citenamefont {Dangi\ifmmode~\acute{c}\else \'{c}\fi{}},
  \citenamefont {Murphy}, \citenamefont {Murray}, \citenamefont {Fahy},\ and\
  \citenamefont {Savi\ifmmode~\acute{c}\else \'{c}\fi{}}}]{OurNTE}%
  \BibitemOpen
  \bibfield  {author} {\bibinfo {author} {\bibfnamefont {{\DJ}.}~\bibnamefont
  {Dangi\ifmmode~\acute{c}\else \'{c}\fi{}}}, \bibinfo {author} {\bibfnamefont
  {A.~R.}\ \bibnamefont {Murphy}}, \bibinfo {author} {\bibfnamefont
  {{\'E}.~D.}\ \bibnamefont {Murray}}, \bibinfo {author} {\bibfnamefont
  {S.}~\bibnamefont {Fahy}},\ and\ \bibinfo {author} {\bibfnamefont
  {I.}~\bibnamefont {Savi\ifmmode~\acute{c}\else \'{c}\fi{}}},\ }\bibfield
  {title} {\bibinfo {title} {Coupling between acoustic and soft transverse
  optical phonons leads to negative thermal expansion of {GeTe} near the
  ferroelectric phase transition},\ }\href
  {https://doi.org/10.1103/PhysRevB.97.224106} {\bibfield  {journal} {\bibinfo
  {journal} {Phys. Rev. B}\ }\textbf {\bibinfo {volume} {97}},\ \bibinfo
  {pages} {224106} (\bibinfo {year} {2018})}\BibitemShut {NoStop}%
\bibitem [{\citenamefont {Dangi{\'{c}}}\ \emph {et~al.}(2021)\citenamefont
  {Dangi{\'{c}}}, \citenamefont {Hellman}, \citenamefont {Fahy},\ and\
  \citenamefont {Savi{\'{c}}}}]{OurTC}%
  \BibitemOpen
  \bibfield  {author} {\bibinfo {author} {\bibfnamefont {{\DJ}.}~\bibnamefont
  {Dangi{\'{c}}}}, \bibinfo {author} {\bibfnamefont {O.}~\bibnamefont
  {Hellman}}, \bibinfo {author} {\bibfnamefont {S.}~\bibnamefont {Fahy}},\ and\
  \bibinfo {author} {\bibfnamefont {I.}~\bibnamefont {Savi{\'{c}}}},\
  }\bibfield  {title} {\bibinfo {title} {The origin of the lattice thermal
  conductivity enhancement at the ferroelectric phase transition in {GeTe}},\
  }\href {https://doi.org/10.1038/s41524-021-00523-7} {\bibfield  {journal}
  {\bibinfo  {journal} {npj Comput. Mater.}\ }\textbf {\bibinfo {volume} {7}},\
  \bibinfo {pages} {57} (\bibinfo {year} {2021})}\BibitemShut {NoStop}%
\bibitem [{\citenamefont {He}\ \emph {et~al.}(2012)\citenamefont {He},
  \citenamefont {Savi{\'c}}, \citenamefont {Donadio},\ and\ \citenamefont
  {Galli}}]{IvanaMD}%
  \BibitemOpen
  \bibfield  {author} {\bibinfo {author} {\bibfnamefont {Y.}~\bibnamefont
  {He}}, \bibinfo {author} {\bibfnamefont {I.}~\bibnamefont {Savi{\'c}}},
  \bibinfo {author} {\bibfnamefont {D.}~\bibnamefont {Donadio}},\ and\ \bibinfo
  {author} {\bibfnamefont {G.}~\bibnamefont {Galli}},\ }\bibfield  {title}
  {\bibinfo {title} {Lattice thermal conductivity of semiconducting bulk
  materials: atomistic simulations},\ }\href
  {https://doi.org/10.1039/C2CP42394D} {\bibfield  {journal} {\bibinfo
  {journal} {Phys. Chem. Chem. Phys.}\ }\textbf {\bibinfo {volume} {14}},\
  \bibinfo {pages} {16209} (\bibinfo {year} {2012})}\BibitemShut {NoStop}%
\bibitem [{\citenamefont {Puligheddu}\ \emph {et~al.}(2019)\citenamefont
  {Puligheddu}, \citenamefont {Xia}, \citenamefont {Chan},\ and\ \citenamefont
  {Galli}}]{GalliMD}%
  \BibitemOpen
  \bibfield  {author} {\bibinfo {author} {\bibfnamefont {M.}~\bibnamefont
  {Puligheddu}}, \bibinfo {author} {\bibfnamefont {Y.}~\bibnamefont {Xia}},
  \bibinfo {author} {\bibfnamefont {M.}~\bibnamefont {Chan}},\ and\ \bibinfo
  {author} {\bibfnamefont {G.}~\bibnamefont {Galli}},\ }\bibfield  {title}
  {\bibinfo {title} {Computational prediction of lattice thermal conductivity:
  A comparison of molecular dynamics and boltzmann transport approaches},\
  }\href {https://doi.org/10.1103/PhysRevMaterials.3.085401} {\bibfield
  {journal} {\bibinfo  {journal} {Phys. Rev. Materials}\ }\textbf {\bibinfo
  {volume} {3}},\ \bibinfo {pages} {085401} (\bibinfo {year}
  {2019})}\BibitemShut {NoStop}%
\bibitem [{\citenamefont {Sangiorgio}\ \emph {et~al.}(2018)\citenamefont
  {Sangiorgio}, \citenamefont {Bozin}, \citenamefont {Malliakas}, \citenamefont
  {Fechner}, \citenamefont {Simonov}, \citenamefont {Kanatzidis}, \citenamefont
  {Billinge}, \citenamefont {Spaldin},\ and\ \citenamefont {Weber}}]{PbTeMD}%
  \BibitemOpen
  \bibfield  {author} {\bibinfo {author} {\bibfnamefont {B.}~\bibnamefont
  {Sangiorgio}}, \bibinfo {author} {\bibfnamefont {E.~S.}\ \bibnamefont
  {Bozin}}, \bibinfo {author} {\bibfnamefont {C.~D.}\ \bibnamefont
  {Malliakas}}, \bibinfo {author} {\bibfnamefont {M.}~\bibnamefont {Fechner}},
  \bibinfo {author} {\bibfnamefont {A.}~\bibnamefont {Simonov}}, \bibinfo
  {author} {\bibfnamefont {M.~G.}\ \bibnamefont {Kanatzidis}}, \bibinfo
  {author} {\bibfnamefont {S.~J.~L.}\ \bibnamefont {Billinge}}, \bibinfo
  {author} {\bibfnamefont {N.~A.}\ \bibnamefont {Spaldin}},\ and\ \bibinfo
  {author} {\bibfnamefont {T.}~\bibnamefont {Weber}},\ }\bibfield  {title}
  {\bibinfo {title} {Correlated local dipoles in pbte},\ }\href
  {https://doi.org/10.1103/PhysRevMaterials.2.085402} {\bibfield  {journal}
  {\bibinfo  {journal} {Phys. Rev. Materials}\ }\textbf {\bibinfo {volume}
  {2}},\ \bibinfo {pages} {085402} (\bibinfo {year} {2018})}\BibitemShut
  {NoStop}%
\bibitem [{\citenamefont {Chatterji}\ \emph {et~al.}(2018)\citenamefont
  {Chatterji}, \citenamefont {Rols},\ and\ \citenamefont
  {Wdowik}}]{ChatterjiMD}%
  \BibitemOpen
  \bibfield  {author} {\bibinfo {author} {\bibfnamefont {T.}~\bibnamefont
  {Chatterji}}, \bibinfo {author} {\bibfnamefont {S.}~\bibnamefont {Rols}},\
  and\ \bibinfo {author} {\bibfnamefont {U.~D.}\ \bibnamefont {Wdowik}},\
  }\bibfield  {title} {\bibinfo {title} {Dynamics of the phase-change material
  {GeTe} across the structural phase transition},\ }\href
  {https://doi.org/10.1007/s11467-018-0864-1} {\bibfield  {journal} {\bibinfo
  {journal} {Front. Phys.}\ }\textbf {\bibinfo {volume} {14}},\ \bibinfo
  {pages} {23601} (\bibinfo {year} {2018})}\BibitemShut {NoStop}%
\bibitem [{\citenamefont {Xu}\ \emph {et~al.}(2018)\citenamefont {Xu},
  \citenamefont {Lei}, \citenamefont {Yuan}, \citenamefont {Xue}, \citenamefont
  {Guo}, \citenamefont {Wang}, \citenamefont {Miao},\ and\ \citenamefont
  {Mazzarello}}]{GeTeAIMD}%
  \BibitemOpen
  \bibfield  {author} {\bibinfo {author} {\bibfnamefont {M.}~\bibnamefont
  {Xu}}, \bibinfo {author} {\bibfnamefont {Z.}~\bibnamefont {Lei}}, \bibinfo
  {author} {\bibfnamefont {J.}~\bibnamefont {Yuan}}, \bibinfo {author}
  {\bibfnamefont {K.}~\bibnamefont {Xue}}, \bibinfo {author} {\bibfnamefont
  {Y.}~\bibnamefont {Guo}}, \bibinfo {author} {\bibfnamefont {S.}~\bibnamefont
  {Wang}}, \bibinfo {author} {\bibfnamefont {X.}~\bibnamefont {Miao}},\ and\
  \bibinfo {author} {\bibfnamefont {R.}~\bibnamefont {Mazzarello}},\ }\bibfield
   {title} {\bibinfo {title} {Structural disorder in the high-temperature cubic
  phase of {GeTe}},\ }\href {https://doi.org/10.1039/C8RA02561D} {\bibfield
  {journal} {\bibinfo  {journal} {RSC Adv.}\ }\textbf {\bibinfo {volume} {8}},\
  \bibinfo {pages} {17435} (\bibinfo {year} {2018})}\BibitemShut {NoStop}%
\bibitem [{\citenamefont {Tersoff}(1989)}]{Tersoff}%
  \BibitemOpen
  \bibfield  {author} {\bibinfo {author} {\bibfnamefont {J.}~\bibnamefont
  {Tersoff}},\ }\bibfield  {title} {\bibinfo {title} {Modeling solid-state
  chemistry: Interatomic potentials for multicomponent systems},\ }\href
  {https://doi.org/10.1103/PhysRevB.39.5566} {\bibfield  {journal} {\bibinfo
  {journal} {Phys. Rev. B}\ }\textbf {\bibinfo {volume} {39}},\ \bibinfo
  {pages} {5566} (\bibinfo {year} {1989})}\BibitemShut {NoStop}%
\bibitem [{\citenamefont {Qiu}\ and\ \citenamefont {Ruan}(2009)}]{RuanBiTe}%
  \BibitemOpen
  \bibfield  {author} {\bibinfo {author} {\bibfnamefont {B.}~\bibnamefont
  {Qiu}}\ and\ \bibinfo {author} {\bibfnamefont {X.}~\bibnamefont {Ruan}},\
  }\bibfield  {title} {\bibinfo {title} {Molecular dynamics simulations of
  lattice thermal conductivity of bismuth telluride using two-body interatomic
  potentials},\ }\href {https://doi.org/10.1103/PhysRevB.80.165203} {\bibfield
  {journal} {\bibinfo  {journal} {Phys. Rev. B}\ }\textbf {\bibinfo {volume}
  {80}},\ \bibinfo {pages} {165203} (\bibinfo {year} {2009})}\BibitemShut
  {NoStop}%
\bibitem [{\citenamefont {Troncoso}\ \emph {et~al.}(2019)\citenamefont
  {Troncoso}, \citenamefont {Aguado-Puente},\ and\ \citenamefont
  {Kohanoff}}]{JavierMD}%
  \BibitemOpen
  \bibfield  {author} {\bibinfo {author} {\bibfnamefont {J.~F.}\ \bibnamefont
  {Troncoso}}, \bibinfo {author} {\bibfnamefont {P.}~\bibnamefont
  {Aguado-Puente}},\ and\ \bibinfo {author} {\bibfnamefont {J.}~\bibnamefont
  {Kohanoff}},\ }\bibfield  {title} {\bibinfo {title} {Effect of intrinsic
  defects on the thermal conductivity of {PbTe} from classical molecular
  dynamics simulations},\ }\href {https://doi.org/10.1088/1361-648x/ab4aa8}
  {\bibfield  {journal} {\bibinfo  {journal} {J. Phys.: Condens. Matter}\
  }\textbf {\bibinfo {volume} {32}},\ \bibinfo {pages} {045701} (\bibinfo
  {year} {2019})}\BibitemShut {NoStop}%
\bibitem [{\citenamefont {Bart\'ok}\ \emph {et~al.}(2010)\citenamefont
  {Bart\'ok}, \citenamefont {Payne}, \citenamefont {Kondor},\ and\
  \citenamefont {Cs\'anyi}}]{GAP1}%
  \BibitemOpen
  \bibfield  {author} {\bibinfo {author} {\bibfnamefont {A.~P.}\ \bibnamefont
  {Bart\'ok}}, \bibinfo {author} {\bibfnamefont {M.~C.}\ \bibnamefont {Payne}},
  \bibinfo {author} {\bibfnamefont {R.}~\bibnamefont {Kondor}},\ and\ \bibinfo
  {author} {\bibfnamefont {G.}~\bibnamefont {Cs\'anyi}},\ }\bibfield  {title}
  {\bibinfo {title} {Gaussian approximation potentials: The accuracy of quantum
  mechanics, without the electrons},\ }\href
  {https://doi.org/10.1103/PhysRevLett.104.136403} {\bibfield  {journal}
  {\bibinfo  {journal} {Phys. Rev. Lett.}\ }\textbf {\bibinfo {volume} {104}},\
  \bibinfo {pages} {136403} (\bibinfo {year} {2010})}\BibitemShut {NoStop}%
\bibitem [{\citenamefont {Bart\'ok}\ \emph {et~al.}(2013)\citenamefont
  {Bart\'ok}, \citenamefont {Kondor},\ and\ \citenamefont {Cs\'anyi}}]{GAP2}%
  \BibitemOpen
  \bibfield  {author} {\bibinfo {author} {\bibfnamefont {A.~P.}\ \bibnamefont
  {Bart\'ok}}, \bibinfo {author} {\bibfnamefont {R.}~\bibnamefont {Kondor}},\
  and\ \bibinfo {author} {\bibfnamefont {G.}~\bibnamefont {Cs\'anyi}},\
  }\bibfield  {title} {\bibinfo {title} {On representing chemical
  environments},\ }\href {https://doi.org/10.1103/PhysRevB.87.184115}
  {\bibfield  {journal} {\bibinfo  {journal} {Phys. Rev. B}\ }\textbf {\bibinfo
  {volume} {87}},\ \bibinfo {pages} {184115} (\bibinfo {year}
  {2013})}\BibitemShut {NoStop}%
\bibitem [{\citenamefont {Behler}\ and\ \citenamefont
  {Parrinello}(2007)}]{NN1}%
  \BibitemOpen
  \bibfield  {author} {\bibinfo {author} {\bibfnamefont {J.}~\bibnamefont
  {Behler}}\ and\ \bibinfo {author} {\bibfnamefont {M.}~\bibnamefont
  {Parrinello}},\ }\bibfield  {title} {\bibinfo {title} {Generalized
  neural-network representation of high-dimensional potential-energy
  surfaces},\ }\href {https://doi.org/10.1103/PhysRevLett.98.146401} {\bibfield
   {journal} {\bibinfo  {journal} {Phys. Rev. Lett.}\ }\textbf {\bibinfo
  {volume} {98}},\ \bibinfo {pages} {146401} (\bibinfo {year}
  {2007})}\BibitemShut {NoStop}%
\bibitem [{\citenamefont {Zuo}\ \emph {et~al.}(2020)\citenamefont {Zuo},
  \citenamefont {Chen}, \citenamefont {Li}, \citenamefont {Deng}, \citenamefont
  {Chen}, \citenamefont {Behler}, \citenamefont {Cs{\'a}nyi}, \citenamefont
  {Shapeev}, \citenamefont {Thompson}, \citenamefont {Wood},\ and\
  \citenamefont {Ong}}]{descriptor1}%
  \BibitemOpen
  \bibfield  {author} {\bibinfo {author} {\bibfnamefont {Y.}~\bibnamefont
  {Zuo}}, \bibinfo {author} {\bibfnamefont {C.}~\bibnamefont {Chen}}, \bibinfo
  {author} {\bibfnamefont {X.}~\bibnamefont {Li}}, \bibinfo {author}
  {\bibfnamefont {Z.}~\bibnamefont {Deng}}, \bibinfo {author} {\bibfnamefont
  {Y.}~\bibnamefont {Chen}}, \bibinfo {author} {\bibfnamefont {J.}~\bibnamefont
  {Behler}}, \bibinfo {author} {\bibfnamefont {G.}~\bibnamefont {Cs{\'a}nyi}},
  \bibinfo {author} {\bibfnamefont {A.~V.}\ \bibnamefont {Shapeev}}, \bibinfo
  {author} {\bibfnamefont {A.~P.}\ \bibnamefont {Thompson}}, \bibinfo {author}
  {\bibfnamefont {M.~A.}\ \bibnamefont {Wood}},\ and\ \bibinfo {author}
  {\bibfnamefont {S.~P.}\ \bibnamefont {Ong}},\ }\bibfield  {title} {\bibinfo
  {title} {Performance and cost assessment of machine learning interatomic
  potentials},\ }\href {https://doi.org/10.1021/acs.jpca.9b08723} {\bibfield
  {journal} {\bibinfo  {journal} {J. Phys. Chem. A}\ }\textbf {\bibinfo
  {volume} {124}},\ \bibinfo {pages} {731} (\bibinfo {year}
  {2020})}\BibitemShut {NoStop}%
\bibitem [{\citenamefont {Bosoni}\ \emph {et~al.}(2019)\citenamefont {Bosoni},
  \citenamefont {Campi}, \citenamefont {Donadio}, \citenamefont {Sosso},
  \citenamefont {Behler},\ and\ \citenamefont {Bernasconi}}]{Bosoni2019}%
  \BibitemOpen
  \bibfield  {author} {\bibinfo {author} {\bibfnamefont {E.}~\bibnamefont
  {Bosoni}}, \bibinfo {author} {\bibfnamefont {D.}~\bibnamefont {Campi}},
  \bibinfo {author} {\bibfnamefont {D.}~\bibnamefont {Donadio}}, \bibinfo
  {author} {\bibfnamefont {G.~C.}\ \bibnamefont {Sosso}}, \bibinfo {author}
  {\bibfnamefont {J.}~\bibnamefont {Behler}},\ and\ \bibinfo {author}
  {\bibfnamefont {M.}~\bibnamefont {Bernasconi}},\ }\bibfield  {title}
  {\bibinfo {title} {Atomistic simulations of thermal conductivity in {GeTe}
  nanowires},\ }\href {https://doi.org/10.1088/1361-6463/ab5478} {\bibfield
  {journal} {\bibinfo  {journal} {J. Phys. D: Appl. Phys.}\
  }\textbf {\bibinfo {volume} {53}},\ \bibinfo {pages} {054001} (\bibinfo
  {year} {2019})}\BibitemShut {NoStop}%
\bibitem [{\citenamefont {Gabardi}\ \emph {et~al.}(2017)\citenamefont
  {Gabardi}, \citenamefont {Baldi}, \citenamefont {Bosoni}, \citenamefont
  {Campi}, \citenamefont {Caravati}, \citenamefont {Sosso}, \citenamefont
  {Behler},\ and\ \citenamefont {Bernasconi}}]{GeTeNN1}%
  \BibitemOpen
  \bibfield  {author} {\bibinfo {author} {\bibfnamefont {S.}~\bibnamefont
  {Gabardi}}, \bibinfo {author} {\bibfnamefont {E.}~\bibnamefont {Baldi}},
  \bibinfo {author} {\bibfnamefont {E.}~\bibnamefont {Bosoni}}, \bibinfo
  {author} {\bibfnamefont {D.}~\bibnamefont {Campi}}, \bibinfo {author}
  {\bibfnamefont {S.}~\bibnamefont {Caravati}}, \bibinfo {author}
  {\bibfnamefont {G.~C.}\ \bibnamefont {Sosso}}, \bibinfo {author}
  {\bibfnamefont {J.}~\bibnamefont {Behler}},\ and\ \bibinfo {author}
  {\bibfnamefont {M.}~\bibnamefont {Bernasconi}},\ }\bibfield  {title}
  {\bibinfo {title} {Atomistic simulations of the crystallization and aging of
  {GeTe} nanowires},\ }\href {https://doi.org/10.1021/acs.jpcc.7b09862}
  {\bibfield  {journal} {\bibinfo  {journal} {J. Phys. Chem. C}\ }\textbf
  {\bibinfo {volume} {121}},\ \bibinfo {pages} {23827} (\bibinfo {year}
  {2017})}\BibitemShut {NoStop}%
\bibitem [{\citenamefont {Sosso}\ \emph {et~al.}(2012)\citenamefont {Sosso},
  \citenamefont {Miceli}, \citenamefont {Caravati}, \citenamefont {Behler},\
  and\ \citenamefont {Bernasconi}}]{GeTeNN2}%
  \BibitemOpen
  \bibfield  {author} {\bibinfo {author} {\bibfnamefont {G.~C.}\ \bibnamefont
  {Sosso}}, \bibinfo {author} {\bibfnamefont {G.}~\bibnamefont {Miceli}},
  \bibinfo {author} {\bibfnamefont {S.}~\bibnamefont {Caravati}}, \bibinfo
  {author} {\bibfnamefont {J.}~\bibnamefont {Behler}},\ and\ \bibinfo {author}
  {\bibfnamefont {M.}~\bibnamefont {Bernasconi}},\ }\bibfield  {title}
  {\bibinfo {title} {Neural network interatomic potential for the phase change
  material {GeTe}},\ }\href {https://doi.org/10.1103/PhysRevB.85.174103}
  {\bibfield  {journal} {\bibinfo  {journal} {Phys. Rev. B}\ }\textbf {\bibinfo
  {volume} {85}},\ \bibinfo {pages} {174103} (\bibinfo {year}
  {2012})}\BibitemShut {NoStop}%
\bibitem [{\citenamefont {Mocanu}\ \emph {et~al.}(2018)\citenamefont {Mocanu},
  \citenamefont {Konstantinou}, \citenamefont {Lee}, \citenamefont {Bernstein},
  \citenamefont {Deringer}, \citenamefont {Cs{\'a}nyi},\ and\ \citenamefont
  {Elliott}}]{MocanuGeTe}%
  \BibitemOpen
  \bibfield  {author} {\bibinfo {author} {\bibfnamefont {F.~C.}\ \bibnamefont
  {Mocanu}}, \bibinfo {author} {\bibfnamefont {K.}~\bibnamefont
  {Konstantinou}}, \bibinfo {author} {\bibfnamefont {T.~H.}\ \bibnamefont
  {Lee}}, \bibinfo {author} {\bibfnamefont {N.}~\bibnamefont {Bernstein}},
  \bibinfo {author} {\bibfnamefont {V.~L.}\ \bibnamefont {Deringer}}, \bibinfo
  {author} {\bibfnamefont {G.}~\bibnamefont {Cs{\'a}nyi}},\ and\ \bibinfo
  {author} {\bibfnamefont {S.~R.}\ \bibnamefont {Elliott}},\ }\bibfield
  {title} {\bibinfo {title} {Modeling the phase-change memory material,
  {Ge$_2$Sb$_2$Te$_5$}, with a machine-learned interatomic potential},\ }\href
  {https://doi.org/10.1021/acs.jpcb.8b06476} {\bibfield  {journal} {\bibinfo
  {journal} {J. Phys. Chem. B}\ }\textbf {\bibinfo {volume} {122}},\ \bibinfo
  {pages} {8998} (\bibinfo {year} {2018})}\BibitemShut {NoStop}%
\bibitem [{\citenamefont {Wang}\ \emph {et~al.}(2021)\citenamefont {Wang},
  \citenamefont {Wu}, \citenamefont {Zeng}, \citenamefont {Embs}, \citenamefont
  {Pei}, \citenamefont {Ma},\ and\ \citenamefont {Chen}}]{Wang2021}%
  \BibitemOpen
  \bibfield  {author} {\bibinfo {author} {\bibfnamefont {C.}~\bibnamefont
  {Wang}}, \bibinfo {author} {\bibfnamefont {J.}~\bibnamefont {Wu}}, \bibinfo
  {author} {\bibfnamefont {Z.}~\bibnamefont {Zeng}}, \bibinfo {author}
  {\bibfnamefont {J.}~\bibnamefont {Embs}}, \bibinfo {author} {\bibfnamefont
  {Y.}~\bibnamefont {Pei}}, \bibinfo {author} {\bibfnamefont {J.}~\bibnamefont
  {Ma}},\ and\ \bibinfo {author} {\bibfnamefont {Y.}~\bibnamefont {Chen}},\
  }\bibfield  {title} {\bibinfo {title} {Soft-mode dynamics in the
  ferroelectric phase transition of {GeTe}},\ }\href
  {https://doi.org/10.1038/s41524-021-00588-4} {\bibfield  {journal} {\bibinfo
  {journal} {npj Comput. Mater.}\ }\textbf {\bibinfo {volume} {7}},\ \bibinfo
  {pages} {118} (\bibinfo {year} {2021})}\BibitemShut {NoStop}%
\bibitem [{\citenamefont {Verdi}\ \emph {et~al.}(2021)\citenamefont {Verdi},
  \citenamefont {Karsai}, \citenamefont {Liu}, \citenamefont {Jinnouchi},\ and\
  \citenamefont {Kresse}}]{Verdi2021}%
  \BibitemOpen
  \bibfield  {author} {\bibinfo {author} {\bibfnamefont {C.}~\bibnamefont
  {Verdi}}, \bibinfo {author} {\bibfnamefont {F.}~\bibnamefont {Karsai}},
  \bibinfo {author} {\bibfnamefont {P.}~\bibnamefont {Liu}}, \bibinfo {author}
  {\bibfnamefont {R.}~\bibnamefont {Jinnouchi}},\ and\ \bibinfo {author}
  {\bibfnamefont {G.}~\bibnamefont {Kresse}},\ }\bibfield  {title} {\bibinfo
  {title} {Thermal transport and phase transitions of zirconia by on-the-fly
  machine-learned interatomic potentials},\ }\href
  {https://doi.org/10.1038/s41524-021-00630-5} {\bibfield  {journal} {\bibinfo
  {journal} {npj Comput. Mater.}\ }\textbf {\bibinfo {volume} {7}},\
  \bibinfo {pages} {156} (\bibinfo {year} {2021})}\BibitemShut {NoStop}%
\bibitem [{\citenamefont {Plimpton}(1995)}]{lammps}%
  \BibitemOpen
  \bibfield  {author} {\bibinfo {author} {\bibfnamefont {S.}~\bibnamefont
  {Plimpton}},\ }\bibfield  {title} {\bibinfo {title} {Fast parallel algorithms
  for short-range molecular dynamics},\ }\href
  {http://www.sciencedirect.com/science/article/pii/S002199918571039X}
  {\bibfield  {journal} {\bibinfo  {journal} {J. Comput. Phys.}\ }\textbf
  {\bibinfo {volume} {117}},\ \bibinfo {pages} {1 } (\bibinfo {year}
  {1995})}\BibitemShut {NoStop}%
    \bibitem{GapGeTe}
\bibinfo{author}{Dangi{\'c}, {\DJ}.}, \bibinfo{author}{Fahy, S.} \&
  \bibinfo{author}{Savi{\'c}, I.}
\newblock \bibinfo{title}{Gaussian approximation potentials (GAP) for germanium telluride}.
\newblock \emph{\bibinfo{journal}{Materials Cloud Archive}} \textbf{\bibinfo{volume}{2021.42}},
   (\bibinfo{year}{2021}).
\bibitem [{\citenamefont {Shinoda}\ \emph {et~al.}(2004)\citenamefont
  {Shinoda}, \citenamefont {Shiga},\ and\ \citenamefont {Mikami}}]{NPT}%
  \BibitemOpen
  \bibfield  {author} {\bibinfo {author} {\bibfnamefont {W.}~\bibnamefont
  {Shinoda}}, \bibinfo {author} {\bibfnamefont {M.}~\bibnamefont {Shiga}},\
  and\ \bibinfo {author} {\bibfnamefont {M.}~\bibnamefont {Mikami}},\
  }\bibfield  {title} {\bibinfo {title} {Rapid estimation of elastic constants
  by molecular dynamics simulation under constant stress},\ }\href
  {https://doi.org/10.1103/PhysRevB.69.134103} {\bibfield  {journal} {\bibinfo
  {journal} {Phys. Rev. B}\ }\textbf {\bibinfo {volume} {69}},\ \bibinfo
  {pages} {134103} (\bibinfo {year} {2004})}\BibitemShut {NoStop}%
\bibitem [{\citenamefont {Tuckerman}\ \emph {et~al.}(2006)\citenamefont
  {Tuckerman}, \citenamefont {Alejandre}, \citenamefont
  {L{\'{o}}pez-Rend{\'{o}}n}, \citenamefont {Jochim},\ and\ \citenamefont
  {Martyna}}]{NVTNPT}%
  \BibitemOpen
  \bibfield  {author} {\bibinfo {author} {\bibfnamefont {M.~E.}\ \bibnamefont
  {Tuckerman}}, \bibinfo {author} {\bibfnamefont {J.}~\bibnamefont
  {Alejandre}}, \bibinfo {author} {\bibfnamefont {R.}~\bibnamefont
  {L{\'{o}}pez-Rend{\'{o}}n}}, \bibinfo {author} {\bibfnamefont {A.~L.}\
  \bibnamefont {Jochim}},\ and\ \bibinfo {author} {\bibfnamefont {G.~J.}\
  \bibnamefont {Martyna}},\ }\bibfield  {title} {\bibinfo {title} {A
  {Liouville}-operator derived measure-preserving integrator for molecular
  dynamics simulations in the isothermal{\textendash}isobaric ensemble},\
  }\href {https://doi.org/10.1088/0305-4470/39/19/s18} {\bibfield  {journal}
  {\bibinfo  {journal} {J. Phys. A: Math. Gen.}\ }\textbf {\bibinfo {volume}
  {39}},\ \bibinfo {pages} {5629} (\bibinfo {year} {2006})}\BibitemShut
  {NoStop}%
    \bibitem{Vesta}Momma, K. \& Izumi, F. \textsc{VESTA3} for three-dimensional visualization of crystal, volumetric and morphology data. {\em J. App. Cryst.} \textbf{44}, 1272-1276 (2011,12)
\bibitem [{sup()}]{supp}%
  \BibitemOpen
  \href@noop {} {}\bibinfo {howpublished} {See Supplemental Material at [url]
  for convergence studies and a more detailed discussion of some effects
  described in the main paper, which includes Refs~\cite{NTEPT1, NTEPT2,
  NTEPT3}.}\BibitemShut {Stop}%
\bibitem [{\citenamefont {Takenaka}(2018)}]{NTEPT1}%
  \BibitemOpen
  \bibfield  {author} {\bibinfo {author} {\bibfnamefont {K.}~\bibnamefont
  {Takenaka}},\ }\bibfield  {title} {\bibinfo {title} {Progress of research in
  negative thermal expansion materials: Paradigm shift in the control of
  thermal expansion},\ }\href {https://doi.org/10.3389/fchem.2018.00267}
  {\bibfield  {journal} {\bibinfo  {journal} {Front. Chem.}\ }\textbf {\bibinfo
  {volume} {6}},\ \bibinfo {pages} {267} (\bibinfo {year} {2018})}\BibitemShut
  {NoStop}%
\bibitem [{\citenamefont {Fang}\ \emph {et~al.}(2015)\citenamefont {Fang},
  \citenamefont {Wang}, \citenamefont {Shang},\ and\ \citenamefont
  {Liu}}]{NTEPT2}%
  \BibitemOpen
  \bibfield  {author} {\bibinfo {author} {\bibfnamefont {H.}~\bibnamefont
  {Fang}}, \bibinfo {author} {\bibfnamefont {Y.}~\bibnamefont {Wang}}, \bibinfo
  {author} {\bibfnamefont {S.}~\bibnamefont {Shang}},\ and\ \bibinfo {author}
  {\bibfnamefont {Z.-K.}\ \bibnamefont {Liu}},\ }\bibfield  {title} {\bibinfo
  {title} {Nature of ferroelectric-paraelectric phase transition and origin of
  negative thermal expansion in {$\mathrm{PbTi}{\mathrm{O}}_{3}$}},\ }\href
  {https://doi.org/10.1103/PhysRevB.91.024104} {\bibfield  {journal} {\bibinfo
  {journal} {Phys. Rev. B}\ }\textbf {\bibinfo {volume} {91}},\ \bibinfo
  {pages} {024104} (\bibinfo {year} {2015})}\BibitemShut {NoStop}%
\bibitem [{\citenamefont {Pan}\ \emph {et~al.}(2019)\citenamefont {Pan},
  \citenamefont {Jiang}, \citenamefont {Nishikubo}, \citenamefont {Sakai},
  \citenamefont {Ishizaki}, \citenamefont {Oka}, \citenamefont {Lin},\ and\
  \citenamefont {Azuma}}]{NTEPT3}%
  \BibitemOpen
  \bibfield  {author} {\bibinfo {author} {\bibfnamefont {Z.}~\bibnamefont
  {Pan}}, \bibinfo {author} {\bibfnamefont {X.}~\bibnamefont {Jiang}}, \bibinfo
  {author} {\bibfnamefont {T.}~\bibnamefont {Nishikubo}}, \bibinfo {author}
  {\bibfnamefont {Y.}~\bibnamefont {Sakai}}, \bibinfo {author} {\bibfnamefont
  {H.}~\bibnamefont {Ishizaki}}, \bibinfo {author} {\bibfnamefont
  {K.}~\bibnamefont {Oka}}, \bibinfo {author} {\bibfnamefont {Z.}~\bibnamefont
  {Lin}},\ and\ \bibinfo {author} {\bibfnamefont {M.}~\bibnamefont {Azuma}},\
  }\bibfield  {title} {\bibinfo {title} {Pronounced negative thermal expansion
  in lead-free {BiCoO$_3$}-based ferroelectrics triggered by the stabilized
  perovskite structure},\ }\href
  {https://doi.org/10.1021/acs.chemmater.9b01969} {\bibfield  {journal}
  {\bibinfo  {journal} {Chem. Mater.}\ }\textbf {\bibinfo {volume} {31}},\
  \bibinfo {pages} {6187} (\bibinfo {year} {2019})}\BibitemShut {NoStop}%
\bibitem [{\citenamefont {Gainza}\ \emph {et~al.}(2020)\citenamefont {Gainza},
  \citenamefont {Serrano-S{\'a}nchez}, \citenamefont {Nemes}, \citenamefont
  {Mart{\'i}nez}, \citenamefont {Fern{\'a}ndez-D{\'i}az},\ and\ \citenamefont
  {Alonso}}]{mdpigete}%
  \BibitemOpen
  \bibfield  {author} {\bibinfo {author} {\bibfnamefont {J.}~\bibnamefont
  {Gainza}}, \bibinfo {author} {\bibfnamefont {F.}~\bibnamefont
  {Serrano-S{\'a}nchez}}, \bibinfo {author} {\bibfnamefont {N.~M.}\
  \bibnamefont {Nemes}}, \bibinfo {author} {\bibfnamefont {J.~L.}\ \bibnamefont
  {Mart{\'i}nez}}, \bibinfo {author} {\bibfnamefont {M.~T.}\ \bibnamefont
  {Fern{\'a}ndez-D{\'i}az}},\ and\ \bibinfo {author} {\bibfnamefont {J.~A.}\
  \bibnamefont {Alonso}},\ }\bibfield  {title} {\bibinfo {title} {Features of
  the high-temperature structural evolution of {GeTe} thermoelectric probed by
  neutron and synchrotron powder diffraction},\ }\href
  {https://www.mdpi.com/2075-4701/10/1/48} {\bibfield  {journal} {\bibinfo
  {journal} {Metals}\ }\textbf {\bibinfo {volume} {10}} (\bibinfo {year}
  {2020})}\BibitemShut {NoStop}%
\bibitem [{\citenamefont {Sist}\ \emph {et~al.}(2018)\citenamefont {Sist},
  \citenamefont {Kasai}, \citenamefont {Hedegaard},\ and\ \citenamefont
  {Iversen}}]{GeTeBo}%
  \BibitemOpen
  \bibfield  {author} {\bibinfo {author} {\bibfnamefont {M.}~\bibnamefont
  {Sist}}, \bibinfo {author} {\bibfnamefont {H.}~\bibnamefont {Kasai}},
  \bibinfo {author} {\bibfnamefont {E.~M.~J.}\ \bibnamefont {Hedegaard}},\ and\
  \bibinfo {author} {\bibfnamefont {B.~B.}\ \bibnamefont {Iversen}},\
  }\bibfield  {title} {\bibinfo {title} {Role of vacancies in the
  high-temperature pseudodisplacive phase transition in {GeTe}},\ }\href
  {https://doi.org/10.1103/PhysRevB.97.094116} {\bibfield  {journal} {\bibinfo
  {journal} {Phys. Rev. B}\ }\textbf {\bibinfo {volume} {97}},\ \bibinfo
  {pages} {094116} (\bibinfo {year} {2018})}\BibitemShut {NoStop}%
\bibitem [{\citenamefont {Shukla}\ \emph {et~al.}(2008)\citenamefont {Shukla},
  \citenamefont {Watanabe}, \citenamefont {Nino}, \citenamefont {Tulenko},\
  and\ \citenamefont {Phillpot}}]{MgO_MD}%
  \BibitemOpen
  \bibfield  {author} {\bibinfo {author} {\bibfnamefont {P.}~\bibnamefont
  {Shukla}}, \bibinfo {author} {\bibfnamefont {T.}~\bibnamefont {Watanabe}},
  \bibinfo {author} {\bibfnamefont {J.}~\bibnamefont {Nino}}, \bibinfo {author}
  {\bibfnamefont {J.}~\bibnamefont {Tulenko}},\ and\ \bibinfo {author}
  {\bibfnamefont {S.}~\bibnamefont {Phillpot}},\ }\bibfield  {title} {\bibinfo
  {title} {Thermal transport properties of {MgO} and {Nd$_2$Zr$_2$O$_7$}
  pyrochlore by molecular dynamics simulation},\ }\href
  {https://www.sciencedirect.com/science/article/pii/S0022311508003632}
  {\bibfield  {journal} {\bibinfo  {journal} {J. Nucl. Mater.}\ }\textbf
  {\bibinfo {volume} {380}},\ \bibinfo {pages} {1} (\bibinfo {year}
  {2008})}\BibitemShut {NoStop}%
\bibitem [{\citenamefont {Hellman}\ \emph {et~al.}(2011)\citenamefont
  {Hellman}, \citenamefont {Abrikosov},\ and\ \citenamefont {Simak}}]{TDEP1}%
  \BibitemOpen
  \bibfield  {author} {\bibinfo {author} {\bibfnamefont {O.}~\bibnamefont
  {Hellman}}, \bibinfo {author} {\bibfnamefont {I.~A.}\ \bibnamefont
  {Abrikosov}},\ and\ \bibinfo {author} {\bibfnamefont {S.~I.}\ \bibnamefont
  {Simak}},\ }\bibfield  {title} {\bibinfo {title} {Lattice dynamics of
  anharmonic solids from first principles},\ }\href
  {https://doi.org/10.1103/PhysRevB.84.180301} {\bibfield  {journal} {\bibinfo
  {journal} {Phys. Rev. B}\ }\textbf {\bibinfo {volume} {84}},\ \bibinfo
  {pages} {180301} (\bibinfo {year} {2011})}\BibitemShut {NoStop}%
\bibitem [{\citenamefont {Hellman}\ and\ \citenamefont
  {Abrikosov}(2013)}]{TDEP2}%
  \BibitemOpen
  \bibfield  {author} {\bibinfo {author} {\bibfnamefont {O.}~\bibnamefont
  {Hellman}}\ and\ \bibinfo {author} {\bibfnamefont {I.~A.}\ \bibnamefont
  {Abrikosov}},\ }\bibfield  {title} {\bibinfo {title} {Temperature-dependent
  effective third-order interatomic force constants from first principles},\
  }\href {https://doi.org/10.1103/PhysRevB.88.144301} {\bibfield  {journal}
  {\bibinfo  {journal} {Phys. Rev. B}\ }\textbf {\bibinfo {volume} {88}},\
  \bibinfo {pages} {144301} (\bibinfo {year} {2013})}\BibitemShut {NoStop}%
\bibitem [{\citenamefont {Hellman}\ \emph {et~al.}(2013)\citenamefont
  {Hellman}, \citenamefont {Steneteg}, \citenamefont {Abrikosov},\ and\
  \citenamefont {Simak}}]{TDEP3}%
  \BibitemOpen
  \bibfield  {author} {\bibinfo {author} {\bibfnamefont {O.}~\bibnamefont
  {Hellman}}, \bibinfo {author} {\bibfnamefont {P.}~\bibnamefont {Steneteg}},
  \bibinfo {author} {\bibfnamefont {I.~A.}\ \bibnamefont {Abrikosov}},\ and\
  \bibinfo {author} {\bibfnamefont {S.~I.}\ \bibnamefont {Simak}},\ }\bibfield
  {title} {\bibinfo {title} {Temperature dependent effective potential method
  for accurate free energy calculations of solids},\ }\href
  {https://doi.org/10.1103/PhysRevB.87.104111} {\bibfield  {journal} {\bibinfo
  {journal} {Phys. Rev. B}\ }\textbf {\bibinfo {volume} {87}},\ \bibinfo
  {pages} {104111} (\bibinfo {year} {2013})}\BibitemShut {NoStop}%
  \bibitem{PbTe1}Kastbjerg, S., Bindzus, N., Søndergaard, M., Johnsen, S., Lock, N., Christensen, M., Takata, M., Spackman, M. \& Brummerstedt Iversen, B. Direct Evidence of Cation Disorder in Thermoelectric Lead Chalcogenides PbTe and PbS. {Adv. Funct. Mater.}. \textbf{23}, 5477-5483 (2013) 
  \bibitem{PbTe2}Christensen, S., Bindzus, N., Sist, M., Takata, M. \& Iversen, B. Structural disorder, anisotropic micro-strain and cation vacancies in thermo-electric lead chalcogenides. { Phys. Chem. Chem. Phys.}. \textbf{18}, 15874-15883 (2016)
  \bibitem{SnTe1}Knox, K., Bozin, E., Malliakas, C., Kanatzidis, M. \& Billinge, S. Local off-centering symmetry breaking in the high-temperature regime of SnTe. { Phys. Rev. B}. \textbf{89}, 014102 (2014)
\end{thebibliography}

\newpage
\onecolumngrid
\appendix
\section{Appendix A: Convergence study with respect to the supercell size}

We show the convergence study with respect to the MD simulation cell size in Supplementary Figure~\ref{suppfig1}. Supplementary Figure~\ref{suppfig1} (a) shows the change of the volume with temperature for 432 (6$\times$6$\times$6), 1024 (8$\times$8$\times$8) and 2000 (10$\times$10$\times$10) atoms cells. We see that the values of the volume are stable for these three supercell sizes. To check the dependence of the critical temperature with different cell sizes, we show the behavior of the order parameter for different temperatures in Supp. Fig.~\ref{suppfig1} (b). We can see that the critical temperature increases with the size of the simulation cell and it does not change significantly between 8$\times$8$\times$8 and 10$\times$10$\times$10 cells. Results for the thermal expansion presented in the main part of the paper were obtained using 10$\times$10$\times$10 cells.

\begin{figure}[h]
\begin{center}
\includegraphics[width = 0.45\textwidth]{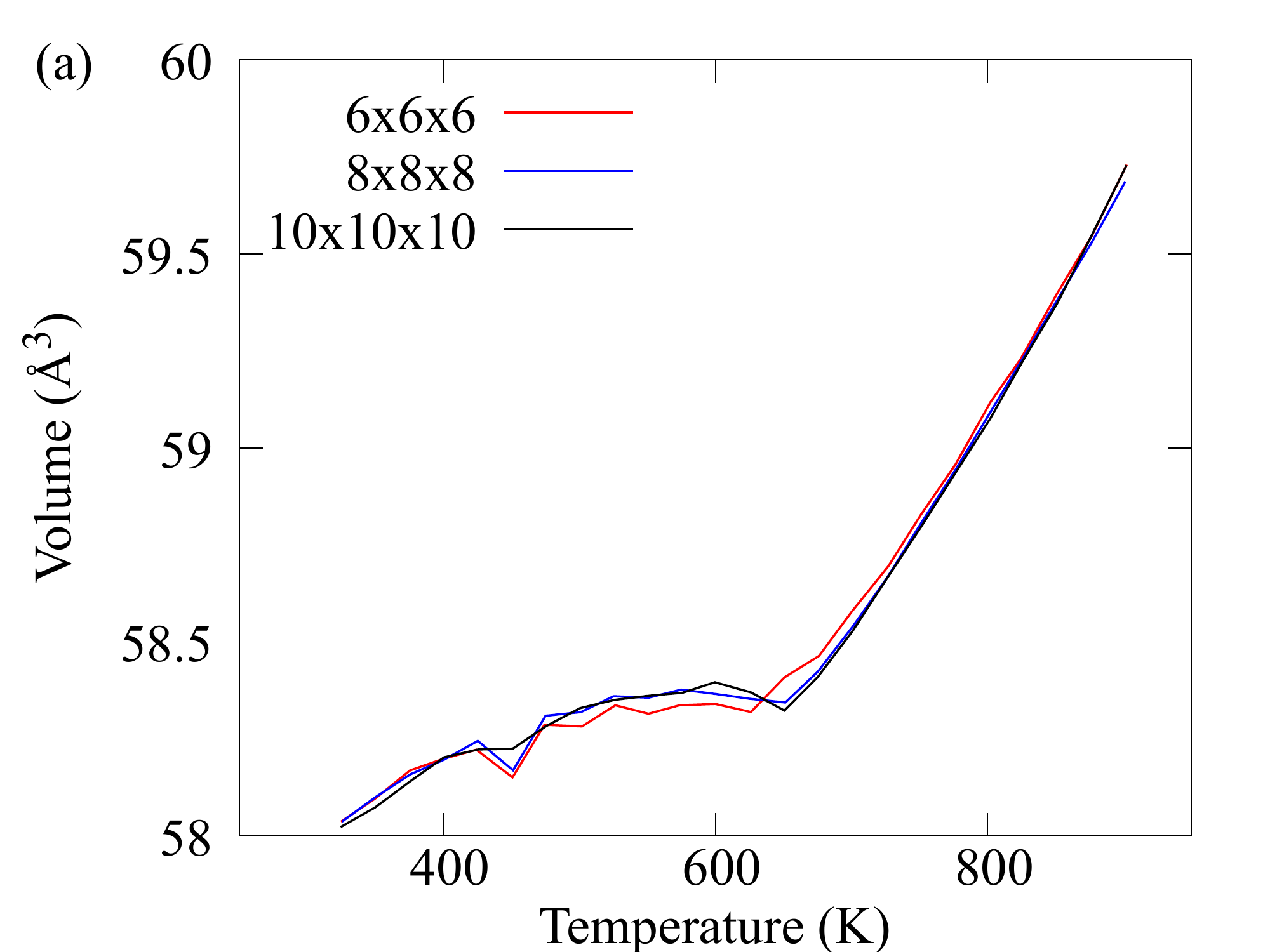}
\includegraphics[width = 0.45\textwidth]{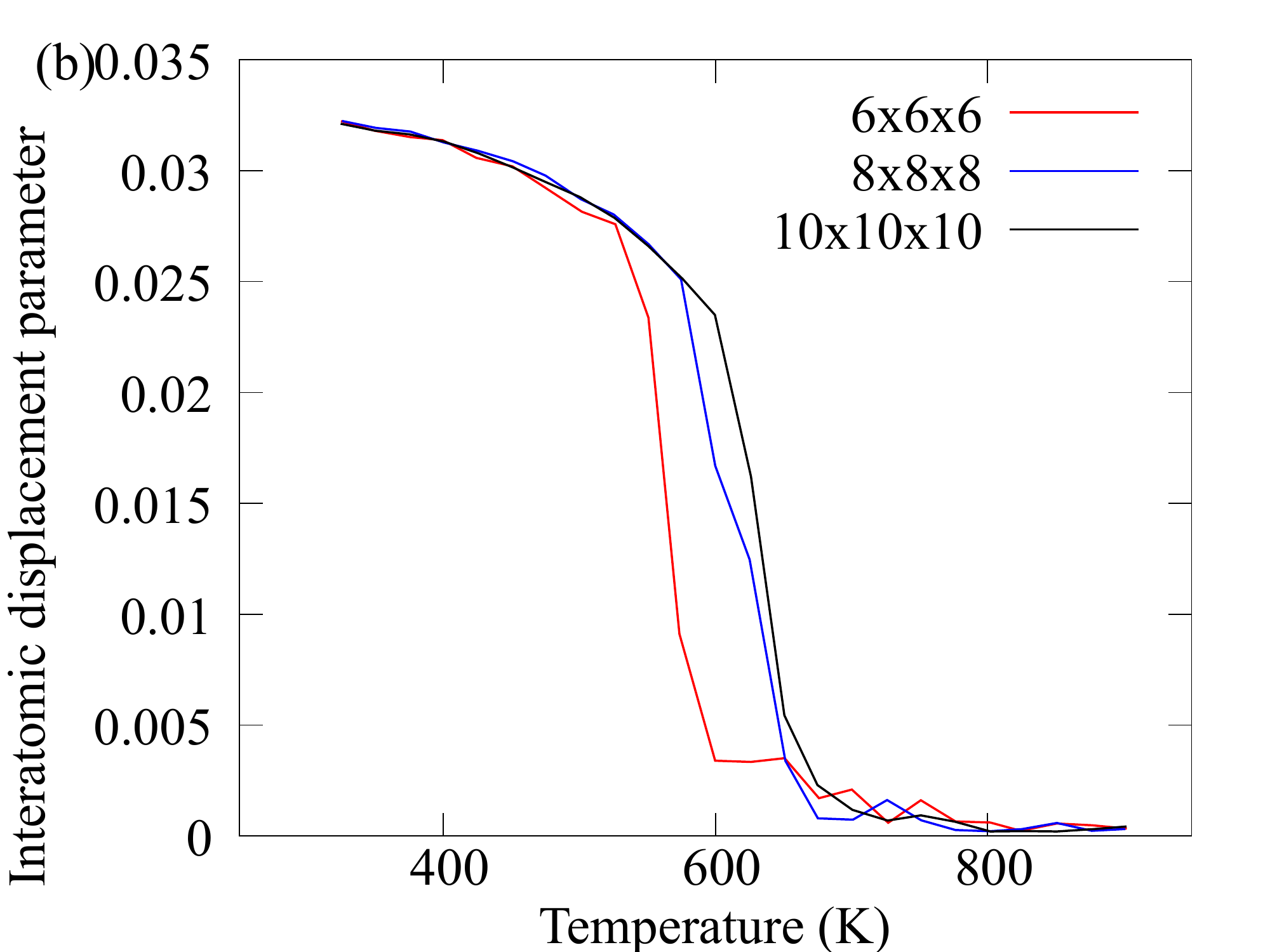}
\end{center}
\caption{The temperature dependence of (a) the volume and (b) the order parameter of germanium telluride for different simulation cell sizes.}
\label{suppfig1}
\end{figure}

\section{Appendix B: Negative thermal expansion at the phase transition}

\begin{figure}
\begin{center}
\includegraphics[width=0.9\linewidth]{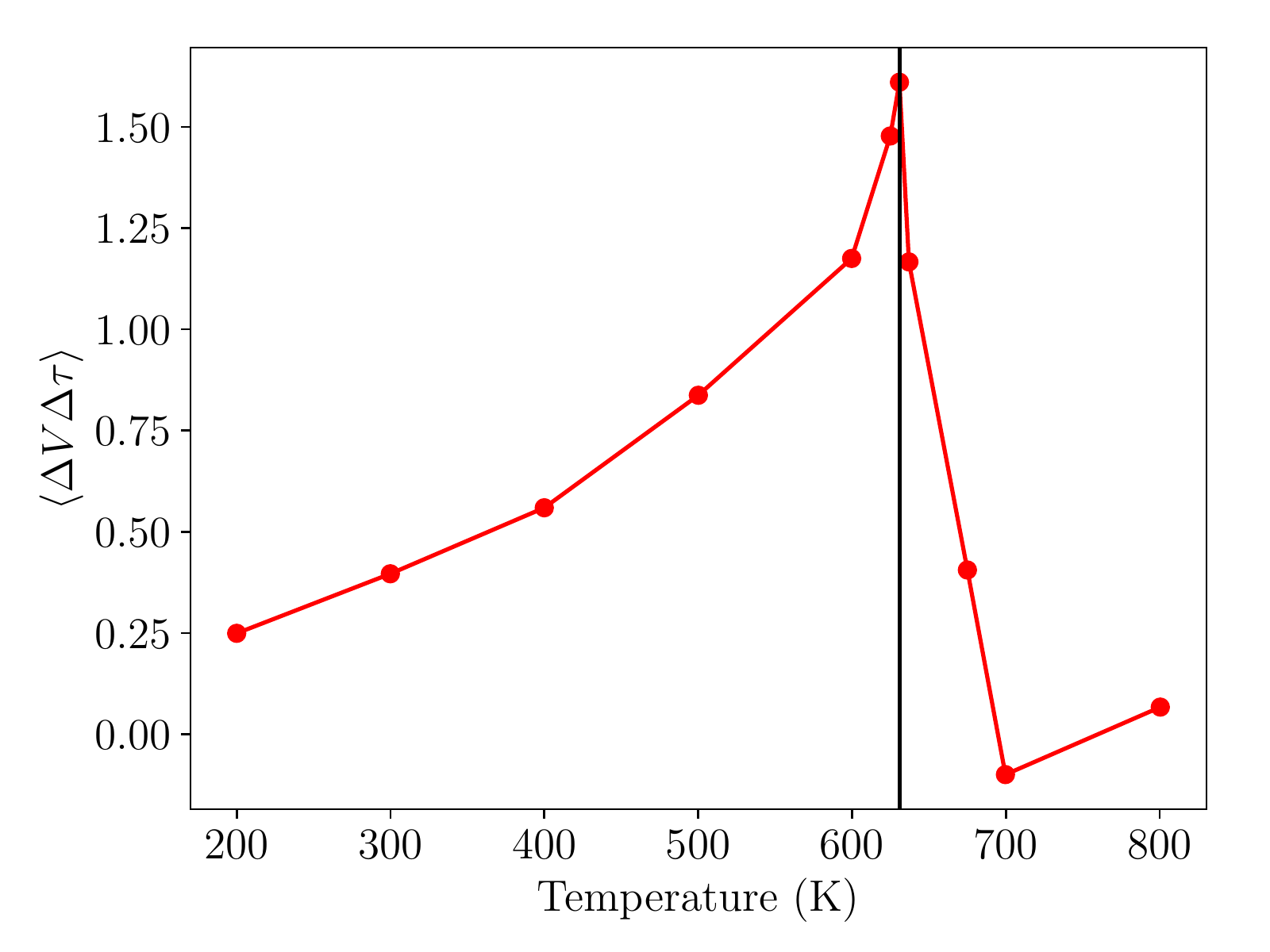}
\caption{Temperature dependence of the strain-order parameter coupling as defined in Eq.~\ref{Strainorderparameter}. The vertical line represents the phase boundary (634 K).}
\label{supfig2}
\end{center}
\end{figure}

Our recent computational study using lattice dynamics showed that the origin of the negative thermal expansion (NTE) at the phase transition in GeTe is an enhancement of acoustic - soft optical mode coupling (or alternatively enhanced strain - order parameter coupling)~\cite{OurNTE}. 
To check this result in molecular dynamics calculations, we can calculate the correlation between the volume and the order parameter:
\begin{align*}
K_{V\tau} = \langle\Delta V\Delta\tau \rangle = \frac{1}{N}\sum ^{N}_{i}(V_i - \langle V\rangle)(\tau _i - \langle \tau\rangle), \numberthis \label{Strainorderparameter}
\end{align*} 
where the $\langle S\rangle$ and $S_i$ are the average and instantaneous values of either the volume $V$ or the order parameter $\tau$, respectively. We show the results for $K_{V\tau}$ for a range of temperatures in Supp. Fig. \ref{supfig2}. This quantity peaks at the phase transition, indicating increased coupling between the volume and the order parameter (enhanced strain - order parameter coupling) and confirming the conclusions of our previous study. 

A recent experimental study claims that negative thermal expansion at the phase transition in GeTe is due to an increase in the number of intrinsic Ge vacancies in the cubic phase~\cite{GeTeBo}. Our study is performed on a stoichiometric GeTe, showing that negative thermal expansion is strongly correlated with the phase transition itself. Negative thermal expansion is observed in a wide range of materials undergoing a phase transition~\cite{NTEPT1, NTEPT2, NTEPT3}. Since these materials have significantly different chemistry and hence different affinities to vacancy formation, NTE in these materials could be correlated with the existence of the phase transition, rather than the vacancy formation. Hence, we speculate that, similarly to GeTe, the strain-order parameter coupling mechanism is also responsible for NTE in these materials.

\section{Appendix C: Radial distribution function for nearest-neighbor bonds of rocksalt structures}

Here we show the radial distribution function for the nearest-neighbor bonds in PbTe and MgO calculated using molecular dynamics simulation at 500 K, see Supplementary Figure~\ref{suppfig5}. We modeled ionic interactions in these materials using already reported interatomic potentials~\cite{JavierMD, MgO_MD}. Both of these compounds are in the rocksalt structure, but their radial distribution functions (RDFs) show similar behavior to GeTe. Both of these RDFs are better fitted using two Gaussian functions. This leads us to believe that the deviation from the pure Gaussian shape of the RDF in GeTe is the consequence of the large anharmonicity of the nearest-neighbor bonds, rather than the evidence of the order-disorder phase transition. This is implicitly confirmed by a larger deviation of the PbTe RDF from the Gaussian shape due to larger anharmonicity of PbTe compared to MgO.

\begin{figure}[h!]
\begin{center}
\includegraphics[width=0.45\textwidth]{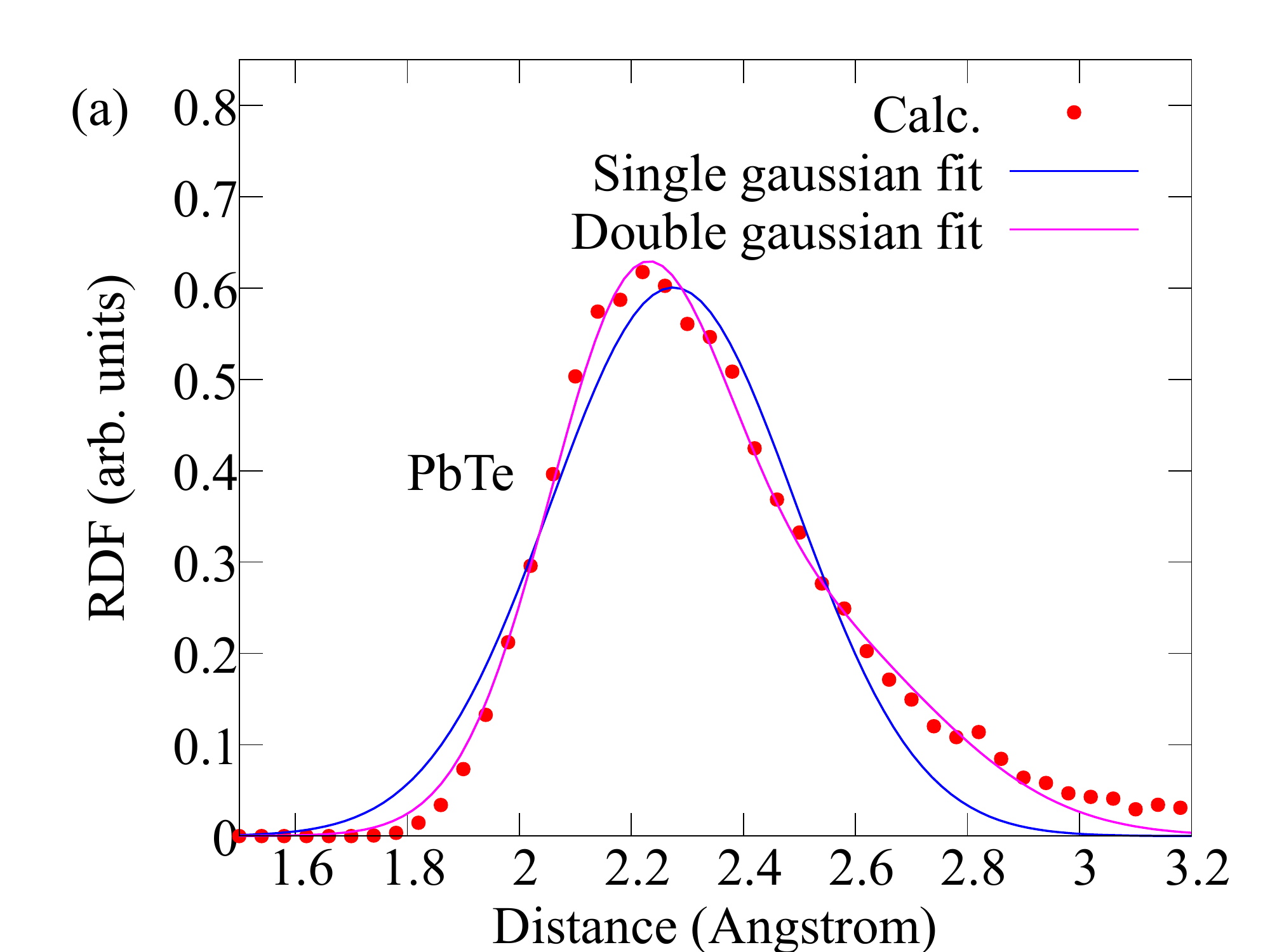}
\includegraphics[width=0.45\textwidth]{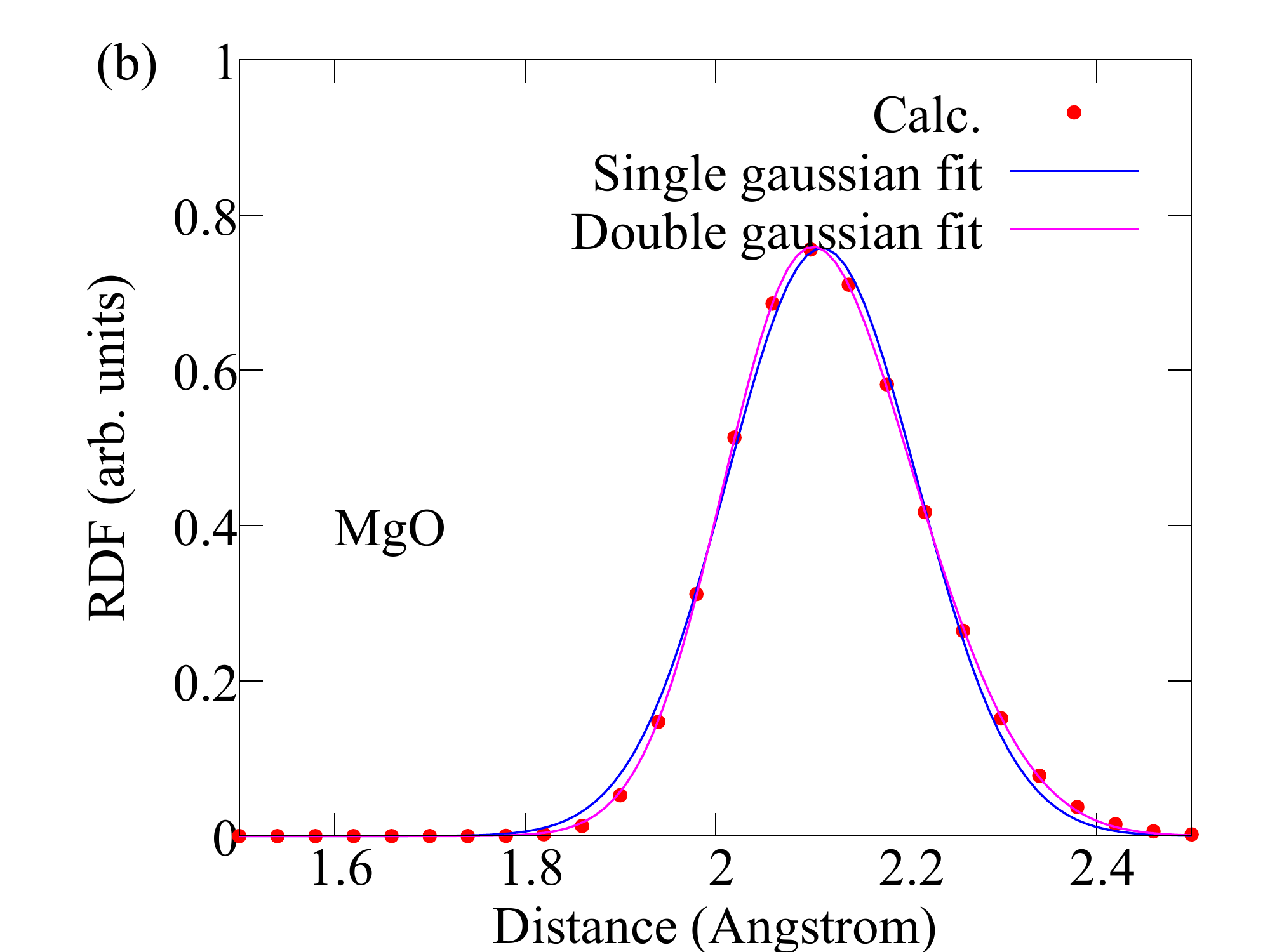}
\caption{Radial distribution function (RDF) of (a) PbTe and (b) MgO at 500 K. The points are our calculations, the blue lines are fits to a single Gaussian, and the magenta lines are fits to two Gaussians.}
\label{suppfig5}
\end{center}
\end{figure}

\begin{figure}[h!]
\begin{center}
\includegraphics[width=0.45\textwidth]{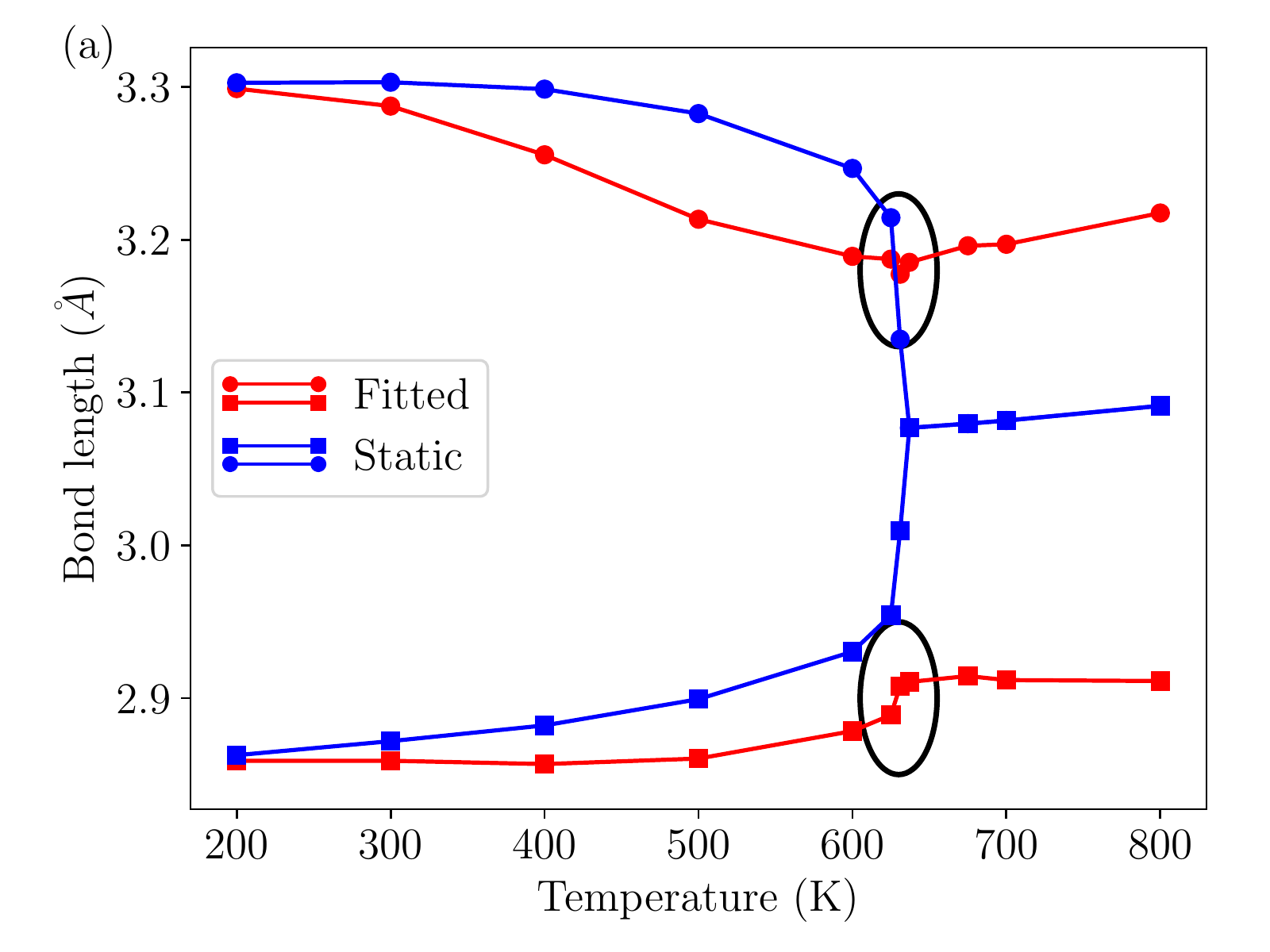}
\includegraphics[width=0.45\textwidth]{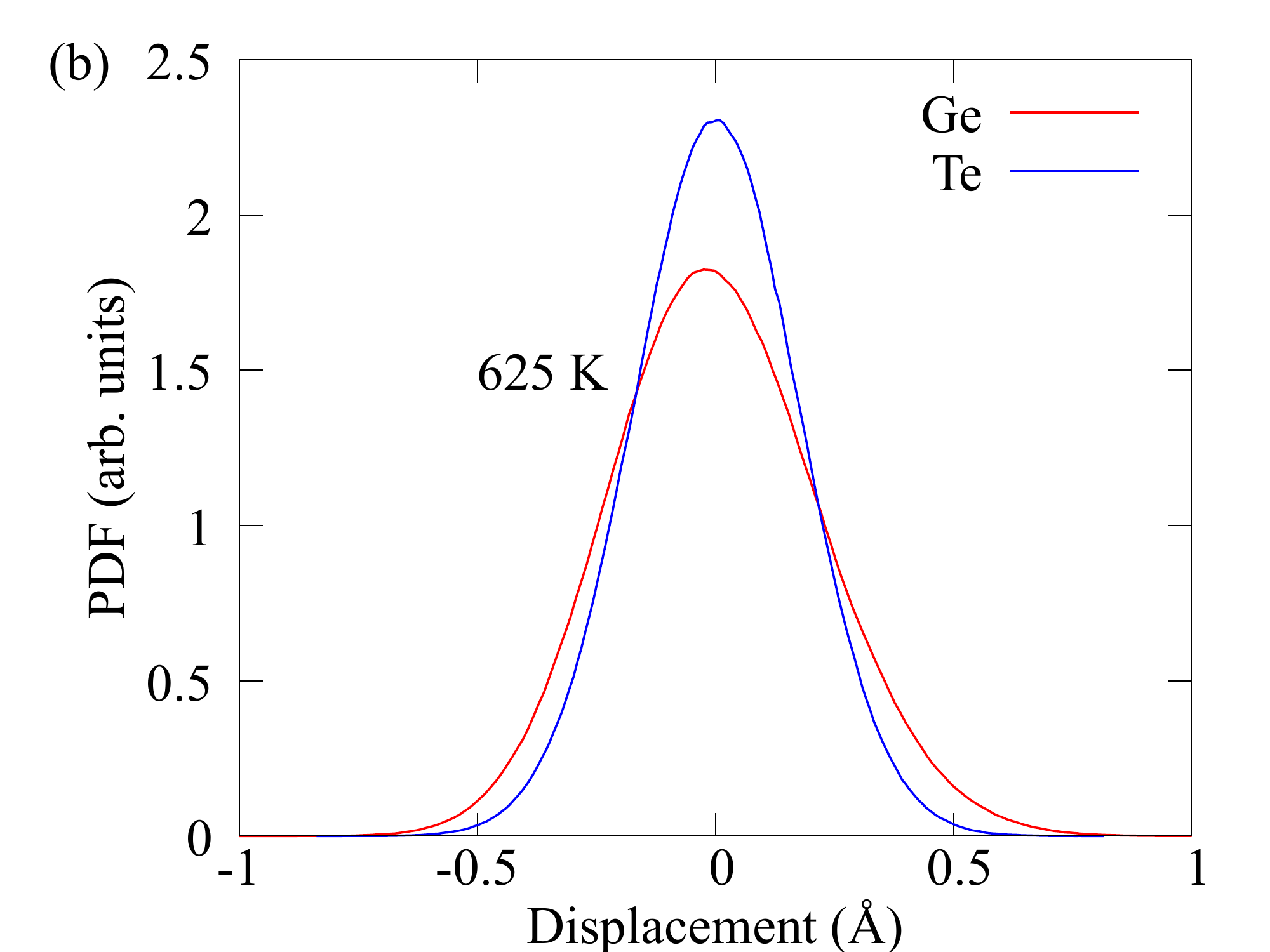}
\caption{(a) The calculated bond lengths using our fitting procedure and the geometry of the unit cell. Black ellipses show the sudden change of the fitted bond lengths at the phase transition. (b) The probability distribution functions of the displacements of Ge and Te atoms along the MD trajectory at 625 K.}
\label{suppfig6}
\end{center}
\end{figure}

Furthermore, we compare the fitted bond lengths versus the "static" bond lengths which are calculated from the geometry of the unit cell of GeTe at that temperature (see Supp. Fig.~\ref{suppfig6} (a)). The nearest-neighbor of a Ge atom is the Te atom located in the nearest-neighbor unit cell. For example, lets say that we look at the primitive unit cell with lattice vector (0,0,0). The first neighbor unit cell would be $-\vec{R}_1$, see Eq. 1 of the main text. The position of the Te atom in Cartesian coordinates in that nearest-neighbor unit cell is then $x_{Te} = a(0,0,3(0.5+\tau) c) - \vec{R}_{1}$, because the position of the Te atom in Cartesian coordinates is $a(0,0,3(\tau + 0.5)c)$. The magnitude of $\vec{x}_{Te}$ vector and thus the bond length is simply: $a\sqrt{b^2 + (3\tau + 0.5)^2c^2}$. This is the longer bond length and we get the shorter bond length if we assume the opposite value of the order parameter for which the position of Te atom in Cartesian coordinates would be: $(0,0,3(0.5 - \tau)c)$. In that case the bond length is: $a\sqrt{b^2+(3\tau - 0.5)^2c^2}$.

We can see that the difference between these two sets of bond lengths increases with temperature. In the cubic phase, the ``static'' nearest-neighbor bond lengths are equal, while the fitted ones are different. However, there is a difference between the fitted and static bond lengths in the rhombohedral phase as well. We checked whether there is a drift of atomic positions during the MD simulation and found that atomic displacements are Gaussians centered around zero for every atom in the MD simulation cell (see Supp. Fig.~\ref{suppfig6} (b)), confirming that the non-Gaussian shape of the bond lengths is not due to the order-disorder character of the phase transition.  

Finally, we would like to draw attention to the noticeable change in the ``fitted'' bond lengths at the phase transition, highlighted with the black ellipses in Supp. Fig.~\ref{suppfig6}. At this temperature we can see a decrease in a larger bond length and an increase in the smaller bond length, a feature that is not obvious from the experimental data. 

\section{Appendix D: Calculation of the order parameter}

Here we will show the procedure we used to compute the order parameter at different temperatures. For each time step, we calculate the local order parameter according to Eq. 3 of the main text. We then select a number of configurations (every 1 ps) and calculate the probability distribution function (PDF) for the local order parameter. We do this by convolving the local order parameter with a Gaussian (the width of the Gaussian is chosen to be two times smaller than the spacing between the sampling points). Finally, we fit these order parameter PDFs and take the fitted value of the mean to be the instantaneous order parameter. The total order parameter is then the arithmetic average over the fitted instantaneous order parameter. This way allows us to better track the most likely value of the order parameter at the phase transition.

\begin{figure}[h!]
\begin{center}
\includegraphics[width=0.6\textwidth]{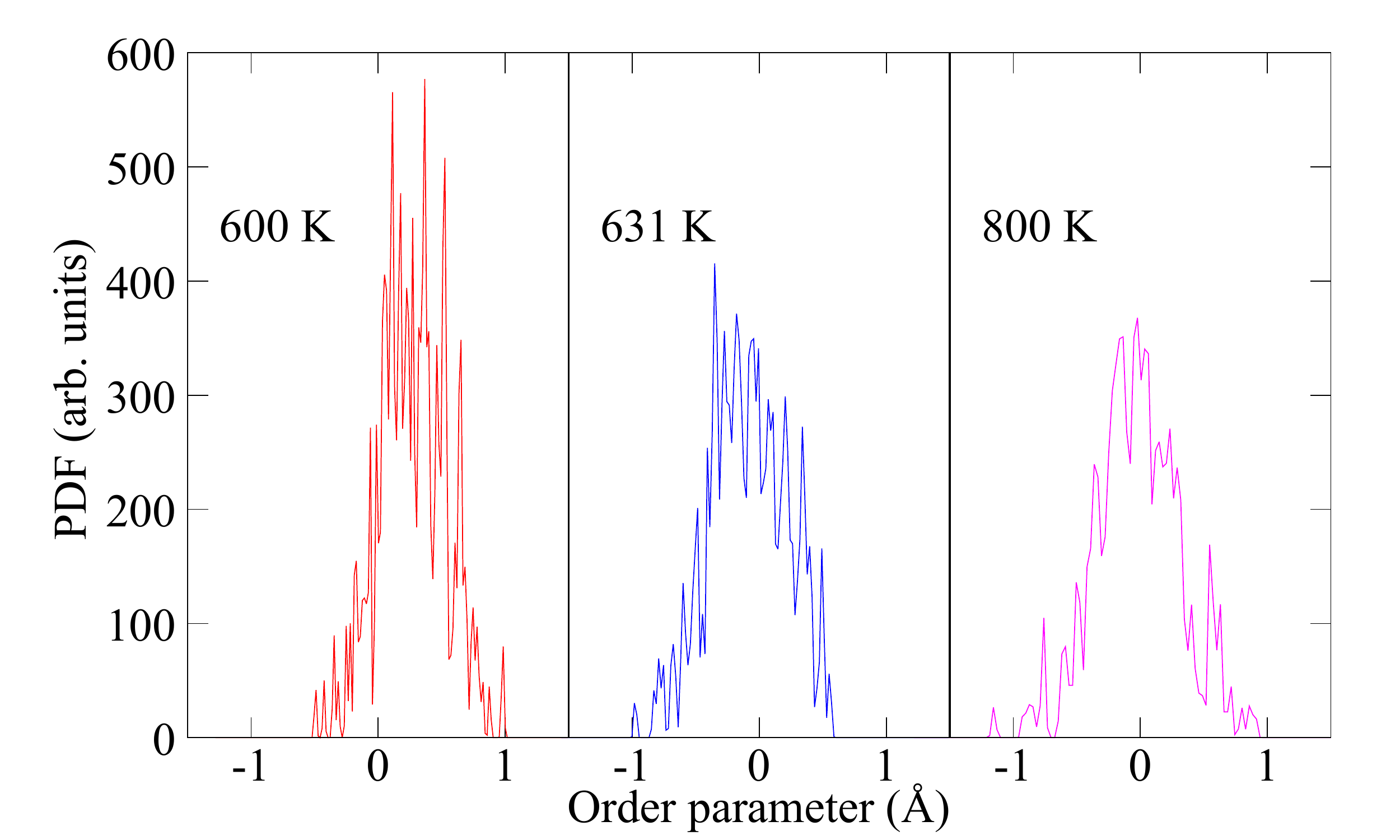}
\caption{The probability distribution functions of the local order parameter at 10 ps for three different temperatures.}
\label{suppfig9}
\end{center}
\end{figure}

The probability distribution functions (PDF) for the local order parameter at three different temperatures and at 10 ps are shown in Supp. Fig.~\ref{suppfig9}. Here we show the order parameter in the units of \r{A}ngstrom, rather than reduced coordinate like in the main part. We fit these PDFs to a Gaussian. The average of the fitted means along an MD trajectory at a certain temperature is the order parameter shown in Fig. 5 (a) of the main part (in the reduced coordinate units). The average of the fitted variance is given in Fig. 5 (b) of the main part. 

\section{Appendix E: Temperature dependence of the order parameter and the soft phonon mode in Landau's model}

The total energy of the system $U$ exhibiting a ferroelectric phase transition is approximated by the double well potential:
\begin{align*}
U(\tau) = A\tau ^2 + B\tau ^4.
\end{align*}
Here $A$ and $B$ are constants and $\tau$ is the order parameter. $A$ is usually taken to be a function of temperature and negative, while $B$ is assumed to be a positive constant. We can find the value of the order parameter at any temperature ($\tau _{0}$) by minimizing total energy of the system:
\begin{align*}
\frac{\partial U}{\partial \tau} = 2A\tau + 4B\tau ^4 = 0  \quad \Rightarrow \quad \tau _{0}(T) = \pm\sqrt{\frac{-A}{2B}}.
\end{align*}
Temperature dependence comes solely through the parameter $A$. In case of the displacive phase transition the order parameter coincides with the eigenvector of the soft phonon mode, which means that the second derivative of the energy with respect to the order parameter is proportional to the square of the frequency of the soft mode $\omega _{0}$.
\begin{align*}
\frac{\partial ^2 U}{\partial \tau^2}|_{\tau _{0}} = 2A + 12B\tau_{0} ^2 = 2A-6A = -4A \sim \omega ^2 _{0}.
\end{align*}
From this analysis we would expect that the soft TO mode and the order parameter have the same $\sqrt{A}$ temperature dependence.

\section{Appendix F: Phonon band structure of GeTe}

\begin{figure}
\begin{center}
\includegraphics[width=0.45\textwidth]{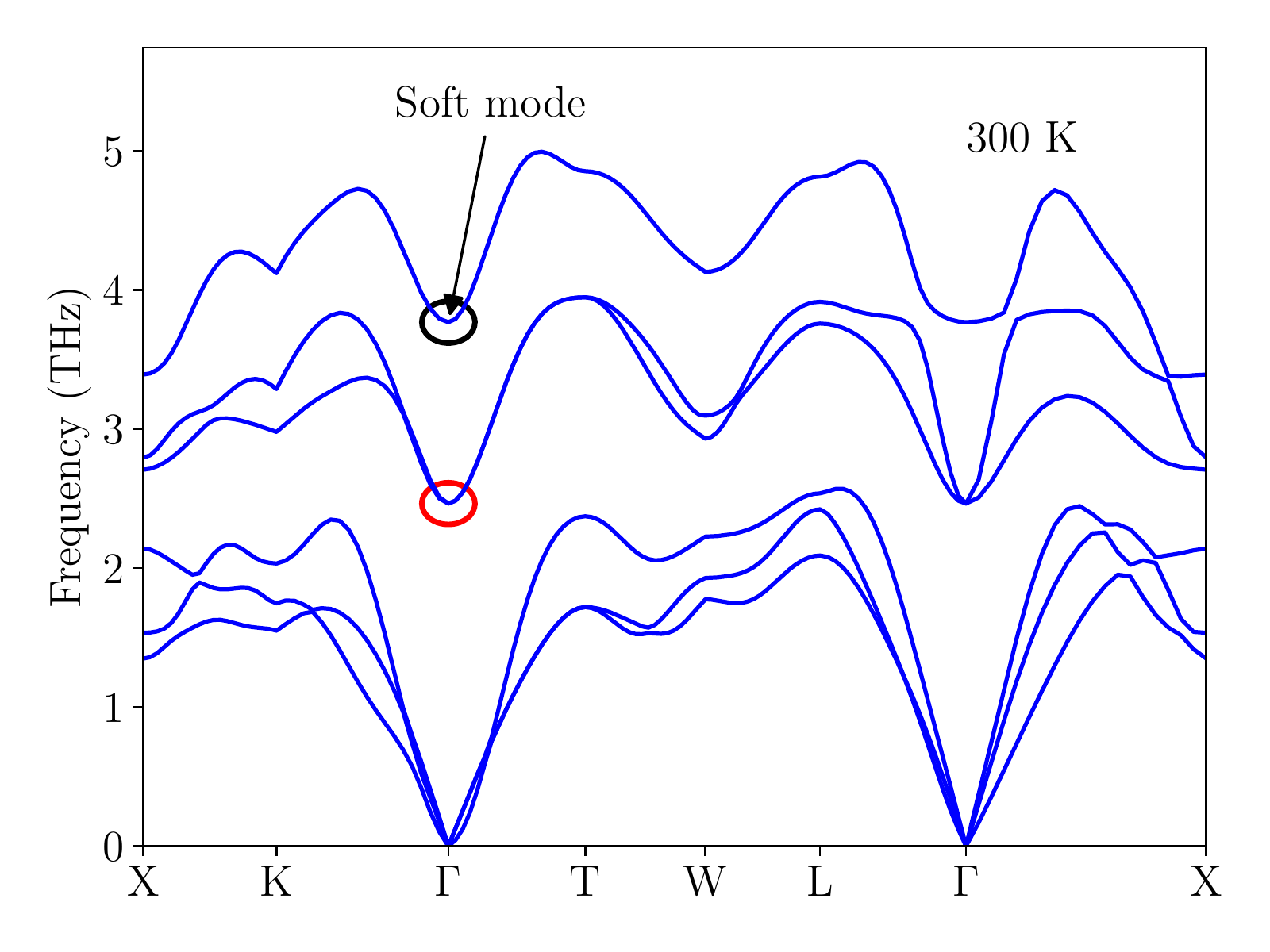}
\caption{Phonon band structure of GeTe at 300 K calculated using the temperature dependent effective potential method. The black ellipse shows the only soft mode in this system (the transverse optical A$_{1g}$ phonon mode). The red ellipse shows other two optical modes (E$_g$ modes) whose frequency corresponds to the oscillations of the correlation functions $\Gamma _{xx} (\vec{q} = 0, \omega)$ and $\Gamma _{yy} (\vec{q} = 0, \omega)$.}
\label{suppfig10}
\end{center}
\end{figure}

In Supp. Fig.~\ref{suppfig10} we show the phonon band structure of GeTe at 300 K calculated using the temperature dependent effective potential method (TDEP)~\cite{TDEP1, TDEP2,TDEP3}. To do this, we collect atomic positions and forces along a molecular dynamics trajectory at 300 K. This calculation is the same calculation we used to extract the order parameter properties. The TDEP method then fits second and third order force constants to these atomic positions and forces. We used the second order force constants to calculate the phonon band structure in Supp. Fig.~\ref{suppfig10}.

\section{Appendix G: Order parameter correlation function}

\begin{figure}[h]
\begin{center}
\includegraphics[width=0.45\textwidth]{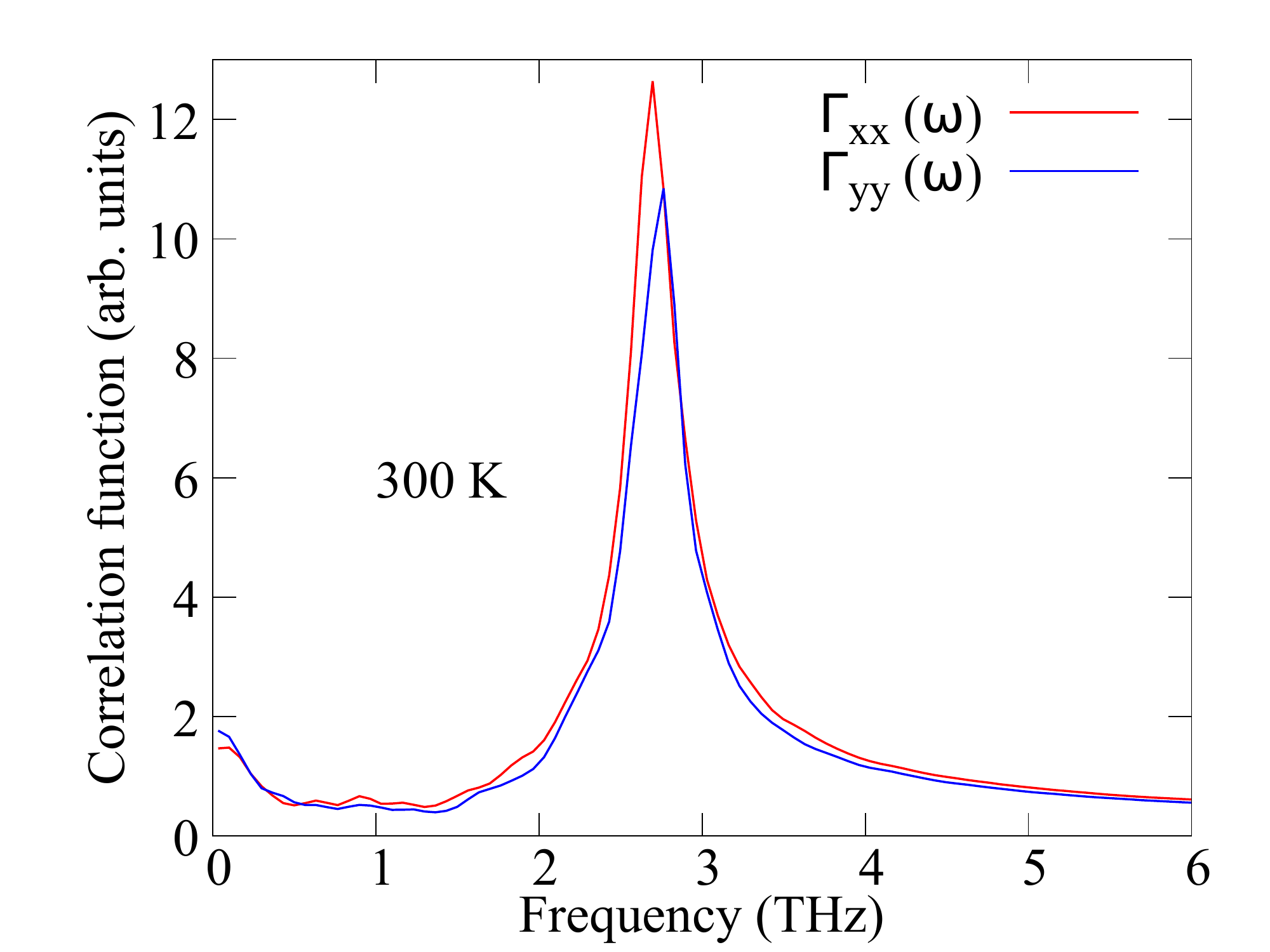}
\caption{The order parameter correlation function in the $x$ and $y$ Cartesian directions at 300 K. }
\label{suppfig7}
\end{center}
\end{figure}

Supplementary Figure~\ref{suppfig7} shows the order parameter correlation function in the $x$ and $y$ Cartesian directions ($\Gamma _{xx} (\vec{q} = 0, \omega)$ and $\Gamma _{yy} (\vec{q} = 0, \omega)$ at 300 K (for definition see Eq. 4 of the main part). The correlation functions peak at the frequency of two other optical modes (two degenerate E$_\text{g}$ modes). The softening of the peak of the correlation function with temperature is present for these two directions as well.

\section{Appendix H: Order parameter correlation length}

We calculated the order parameter correlation length at different temperatures. The procedure is as follows. First, we calculate $G_{zz}(\vec{r}, t = 0)$ for different temperatures (see Eq. 4 of the main part for the definition of $G_{zz}(\vec{r}, t = 0)$). We then choose the largest correlation for a specific distance $r = |\vec{r}|$ and fit $G^{max}_{zz}(\vec{r}, t = 0)$ to a decaying exponential. We have found that the fit is much better if we exclude the $r = 0$ point. The results presented here are obtained without including the $r = 0$ point in the fit. Including this point leads to the same results qualitatively, only with a smaller value of the correlation length at the phase transition.

\begin{figure}[h]
\begin{center}
\includegraphics[width=0.45\textwidth]{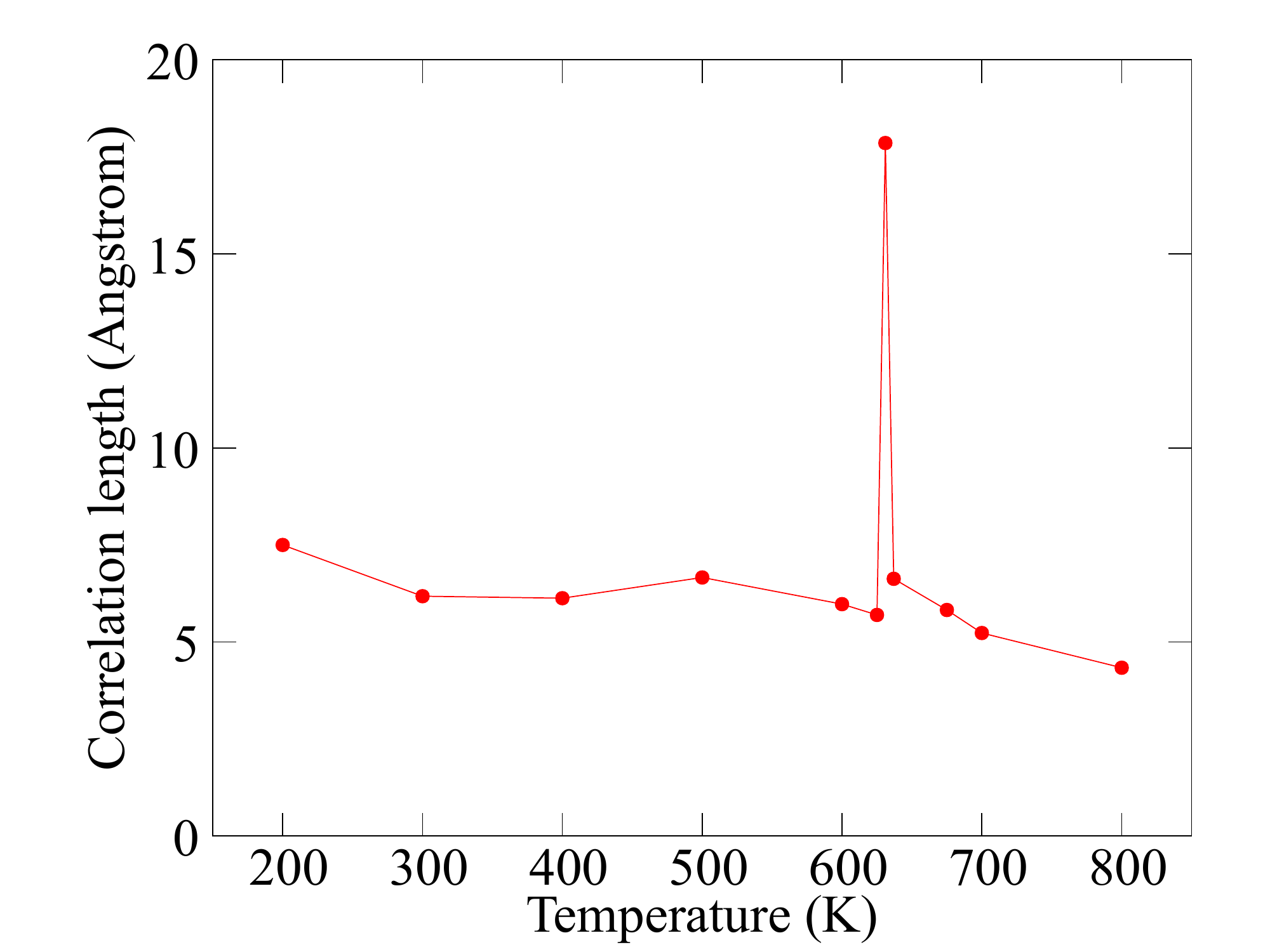}
\caption{The order parameter correlation length of GeTe at different temperatures. }
\label{suppfig8}
\end{center}
\end{figure}

Supplementary Figure~\ref{suppfig8} shows the results of this study. We can see a large increase in the correlation length at the phase transition. However, we can not see whether this divergence follows a specific power law. The correlation length does not change much away from the transition temperature. This behavior is probably the consequence of the finite size of the simulation cell.

\end{document}